\documentstyle[eqsecnum,aps,prd,epsfig]{revtex}
\date{\today}
\tighten
\begin{document}
\tighten
%
%
\def\ltsima{$\; \buildrel < \over \sim \;$}
\def\simlt{\lower.5ex\hbox{\ltsima}}
\def\gtsima{$\; \buildrel > \over \sim \;$}
\def\simgt{\lower.5ex\hbox{\gtsima}}
\newcommand{\Ms}{M_{\odot}}
\newcommand{\gras}[1]{\mbox{\boldmath $#1$}}
\newcommand{\m}{\langle}
\newcommand{\M}{\rangle}
\def \be {\begin{equation}}
\def \ee {\end{equation}}
\def \bea{\begin{eqnarray}}
\def \eea{\end{eqnarray}}
\def \a {{\alpha}}
\def \b {\beta}
\def \r {{\bf r}}
\def \n {\bf \hat{n}}
\def \dphi {\Delta \Phi}
\def \phid {\phi_D}
\def \tPhi {\tilde \Phi}
\def \Ck {{\cal C}_k}
\def \Sk {{\cal S}_k}
\def \L {\gras {\Lambda}}
\def \l {\gras {\lambda}}
\def \bdl {\Delta \gras{\lambda}}
\def \bdL {\Delta \gras{\Lambda}}
\def \dl {\Delta \lambda}
\def \dL {\Delta \Lambda} 
\def \h {{1 \over 2}}
\def \Vp {V_{\cal P}}
\def \tVp {\tilde V_{\cal P}}
\def \tg {{\tilde g}}
\def \tgamma {\tilde \gamma}
\def \tgm {\tilde \gamma}
\def \fM {f_{\rm max}}
\def \cT {{\cal T}}
\def \AM {A_{\rm max}}
\def \Am {A_{\rm min}}
\def \OM {\Omega_{\rm max}}
\def \Om {\Omega_{\rm min}}
\def \apM {a_{\rm p,max}}
\def \apm {a_{\rm p,min}}
\def \aM {a_{\rm max}}
\def \am {a_{\rm min}}
\def \oM {\omega_{\rm max}}
\def \om {\omega_{\rm min}}
\def \alM {\alpha_{\rm max}}
\def \alm {\alpha_{\rm min}}
\def \psM {\psi_{\rm max}}
\def \psm {\psi_{\rm min}}
\def \S {{\cal S}}
%
%
\title{Searching for continuous gravitational wave sources in binary systems}
\author{Sanjeev V. Dhurandhar}
\address{Inter-University Centre for Astronomy and Astrophysics \\
Post Bag 4, Ganeshkhind, Pune 411007, India \\
Max Planck Institut f\"{u}r Gravitationsphysik, 
Albert Einstein Institut\\
Am M\"{u}hlenberg 5, D-14476 Golm, Germany\\
Electronic address: sdh@iucaa.ernet.in}
\author{Alberto Vecchio} 
\address{Max Planck Institut f\"{u}r Gravitationsphysik, 
Albert Einstein Institut\\
Am M\"{u}hlenberg 5, D-14476 Golm, Germany\\
Electronic address: vecchio@aei-potsdam.mpg.de}

\maketitle

\begin{abstract}
We consider the problem of searching for continuous gravitational wave (CW) 
sources orbiting  a companion object. This issue is of particular
interest because, the Low Mass X-ray Binaries (LMXB's), and among them 
Sco X-1 -- the brightest X-ray source in the sky -- might be marginally detectable 
with $\approx 2$ years coherent observation time by the Earth-based laser 
interferometers expected to come on line by 2002, and clearly observable
by the second generation of detectors. Moreover, several radio pulsars,
which could be deemed to be CW sources, are found to orbit 
a companion star or planet, and the LIGO/VIRGO/GEO600 network plans to
continuously monitor such systems.

We estimate the computational costs for a search launched over the additional 
five parameters describing generic
elliptical orbits (up to $e\simlt 0.8$) using match filtering techniques. These 
techniques provide the optimal signal-to-noise ratio and
also a very clear and transparent theoretical framework. Since matched 
filtering will be  
implemented in the final and the most computationally expensive stage 
of the hierarchical strategies, the theoretical framework provided here can be 
used to determine the computational costs. In order to disentangle the 
computational burden
involved in the orbital motion of the CW source, from the other source parameters
(position in the sky and spin-down), and reduce the complexity of the
analysis, we assume that the source is monochromatic (there is no intrinsic 
change in its frequency) and its location in the sky is exactly known. 
The orbital elements, on the other hand, are either assumed to be completely 
unknown or only partly known.

We provide ready-to-use analytical expressions for the number
of templates required to carry out the searches in the astrophysically 
relevant regions of the parameter space, and how the computational cost 
scales with the ranges of the parameters. 
We also determine the critical accuracy to which a particular parameter must 
be known, so that no search is needed for it; we provide rigorous 
statements, based on the  
geometrical formulation of data analysis, concerning the size
of the parameter space so that a particular neutron star is a one-filter target.
This result is formulated in a completely general form, independent
of the particular kind of source, and can be applied to any class of signals 
whose waveform can be accurately predicted.

We apply our theoretical analysis to Sco X-1 and the 44 neutron stars with
binary companions which are listed in the most updated version of the radio
pulsar catalogue. For up to $\approx 3$ hours of coherent integration time, 
Sco X-1 will need at most a few templates; for a week's integration time 
the number of templates rapidly rises to $\simeq 5\times 10^6$.
This is due to the rather poor measurements available today of the projected 
semi-major axis and the orbital phase of the neutron star. 
If, however, the same search is to be carried out with only a few filters, 
then more refined measurements of the orbital parameters are called for - 
an improvement of about three orders of magnitude in the accuracy
is required. Further, we show that the five NS's
(radio pulsars) for which the upper-limits on the signal
strength are highest, require no more than a few templates each
and can be targeted very cheaply in terms of CPU time.

Blind searches of the parameter space of orbital elements are, in general,
completely un-affordable for present or near future dedicated computational
resources, when the coherent integration time is of the order of
the orbital period or longer. For wide binary systems, when the
observation covers only a fraction of one orbit, the computational
burden reduces enormously, and becomes affordable for a significant
region of the parameter space.
\end{abstract}
\pacs{Pacs Number(s): 04.80.Nn, 95.55.Ym, 95.75.Pq,97.60.Gb}

\section{Introduction}
\label{sec:intro}

The construction of several large-scale interferometric
gravitational wave detectors, with optimal sensitivity in the
frequency window $\sim$ 10 Hz -- 1 kHz is close to completion.  
For a period extending upto a year, starting from the end of year 2000, 
engineering data runs will be carried out in order to test and debug the detector 
components.  Finally, between 2002 and 2004 the interferometers
will carry out the first set of "science" data-runs 
with the {\it realistic goal of  directly observing 
gravitational waves} (GW's). This initial phase will be followed by 
substantial upgrades on most of the instruments aimed at reaching  better 
sensitivity and  larger observational bands. The projects include the 
Laser Interferometer Gravitational-wave Observatory (LIGO) in the USA
consisting of two facilities, one at Hanford (WA) and the other at 
Livingston (LA), hosting two 4-km and one 2-km 
interferometers~\cite{ligo}; VIRGO, a French/Italian project located 
at Cascina, near Pisa, Italy, which consists of a 3-km detector (and currently
running about one-to-two years behind the time-frame presented above)
~\cite{virgo}; GEO600, a German/British 0.6-km interferometer located 
at Ruthe, near Hannover, Germany; in Japan, the TAMA project is 
{\it currently running} a medium scale
interferometric detector of arm length 300 meters, and is planning
to extend the baseline to 3 km and carry out other substantial improvements
on the instrument within the next few years~\cite{tama}; finally, the 
ACIGA consortium will build, if funding is approved, a 500 meter interferometer
(AIGO500) near Perth, Australia~\cite{aigo}. In the meantime,
a number of existing resonant bar detectors are steadily increasing their
sensitivity in narrow bandwidths ($\sim 1$ Hz) covering the kHz spectral 
window~\cite{Cerdonio} and serious efforts are being made to fly, 
possibly by 2010, a space-borne laser interferometer~\cite{lisa} 
(LISA: the Laser Interferometer Space Antenna).
LISA will open the low frequency window $10^{-5}$ Hz -- $10^{-2}$ Hz, currently
accessible only via the Doppler tracking of interplanetary
space-craft~\cite{BVI99}, and is projected to have sensitivity several orders of 
magnitude better.

Building and running such powerful machines represents an enormous enterprise,
which has started over 30 years ago. However, analyzing the large amount of
data -- several Mbytes per second -- and digging out with high confidence 
astrophysical signals from the noise which severely corrupts the data, 
presents its own challenges. In fact, over the past few years, interest has been
growing in the area of GW data analysis, which is now regarded as one of the
key aspects for the successful detection of GWs.

Several types of GW sources have been envisaged which could be directly 
observed by Earth-based detectors 
(see~\cite{Thorne87,Thorne95,Fl_GR15,Schutz99} and 
references therein for recent reviews): 
(i) Burst sources -- such as binary systems of neutron stars (NS) and/or 
black holes (BH) in their in-spiral phase, BH/BH and/or BH/NS mergers, and 
supernovae explosions -- whose signals last for a time much shorter, 
typically between a few milli-seconds and a few minutes, than
the planned observational time; (ii) stochastic backgrounds of radiation,
either of primordial or astrophysical origin, and (iii) continuous wave (CW) sources
-- e.g. rapidly rotating neutron stars -- 
where a weak deterministic signal is continuously present in the data stream.

In this paper we deal with data analysis issues regarding CW sources.
More specifically, we investigate the computational load involved in 
filtering the data stream to search for {\it monochromatic
radiation emitted by a neutron star orbiting a companion object}.

\subsection{The data analysis challenge}
\label{subs:dan}

CW emitters pose one of the most computationally intensive problems in
GW data analysis. The weakness of the expected signal requires
very long observation times, of the order of a year (or possibly more)
for accumulating enough signal-to-noise ratio (SNR) so that detection 
is possible. During the observation time, a monochromatic signal in the 
source reference frame, is Doppler modulated by the motion of the detector 
carried by the spinning Earth orbiting the Sun. 
The emitted energy is spread over $\simeq 2\times 10^6\,(T/10^7\,{\rm sec})^2\,
(f/1\,{\rm kHz})$ frequency bins of width $\Delta f = 1/T$, where $T$ is the
time of observation (the formula holds for $T$ upto six months, after that 
the number of bins increases linearly with $T$). In order to recover the whole 
power in one frequency bin one has to "correct" the recorded data stream 
for each possible source position in the sky.
The problem is made worse, if the the intrinsic frequency of the source changes, 
say due to spindown. Then the power is spread over 
$3\times 10^6\,(\tau/10^3\,{\rm yrs})^{-1}\,(T/10^7\,{\rm sec})^2\,(f/1\,{\rm kHz})$
bins, where $\tau = f/\dot{f}$ is (twice) the spin-down age of the NS. Indeed,
one then needs to correct, in addition, for this effect searching through one or 
more spin-down parameters. It is clear that searches for CW's
are limited by the available computational resources~\cite{Schutz89,BCCS98}. 
The optimal technique of tracking coherently the signal phase -- matched
filtering -- is not affordable for integration times longer than $\sim$ 1 to 10
days (the integration time depends on the range of parameters over which the 
search is being carried out). So far, the only viable strategy for a "blind" 
search over a wide range 
of parameters is based on a hierarchical structure~\cite{BC00,SP99}. In this type 
of search, coherent and incoherent search stages are alternated in order to 
identify candidate signals -- and the relevant parameter range -- "cheaply"
by a suboptimal algorithm. Then, the candidate signals are followed-up 
with a coherent and computationally expensive search over the entire observation time 
-- this step constitutes the bottle neck for processing power -- 
to recover the optimal SNR.

However, due to the large computational burden, the algorithms investigated 
so far have been restricted to {\it isolated} NS's. The problem of searching for
a CW emitter orbiting a companion object is simply considered computationally
intractable, as it would require upto five more search parameters which would 
compound the already enormous computational cost. 

To gain insight into the additional Doppler modulation introduced on the 
signal phase by the source motion around a companion, let us consider 
a NS, of mass $m_{\rm NS} = 1.4\Ms$,
with an orbiting companion object of mass $m_2$ and period $P$. During
the observation time $T$, the NS velocity changes by 
$\Delta v = (2\pi/P)^{4/3}\,(G m_{\rm NS})^{1/3}\,q/(1+q)^{2/3}\,T$, where
$q \equiv m_2/m_{\rm NS}$. The frequency is Doppler shifted by 
$\Delta_D\,f = \pm f \Delta v/c$ which has to be
compared to the frequency resolution  $\Delta f$. 
(Here and after, $G$ denotes the Newton's gravitational constant and $c$ denotes 
the speed of light or actually GW).  
The maximum observational time $T_{{\rm max}}$ after which one
needs to correct for the CW's source orbital motion is therefore,
\be
T_{{\rm max}} \simeq 131.6 \, \left[\frac{(1+q)^{1/3}}{\sqrt{q}}\right]\,
\left(\frac{P}{1\,{\rm day}}\right)^{2/3}\,
\left(\frac{m_{NS}}{1.4\,\Ms}\right)^{-1/6}\,
\left(\frac{f}{1\,{\rm kHz}}\right)^{-1/2}\,{\rm sec}\,.
\label{Tmax}
\ee
Of course, this time scale can vary
by orders of magnitude, depending on the NS companion and the orbital
period. For the relevant astrophysical situations we have:
\be
T_{{\rm max}} \simeq
\left\{ \begin{array}{ll}
1.2\times 10^5\,\left(\frac{P}{10^7\,{\rm sec}}\right)^{2/3}\,
\left(\frac{m_{2}}{10^{-3}\,\Ms}\right)^{-1/2}\,
\left(\frac{f}{1\,{\rm kHz}}\right)^{-1/2}\,{\rm sec}
& (m_2 \ll m_{\rm NS}) \nonumber\\
1.0\times 10^2\,\left(\frac{P}{12\,{\rm hours}}\right)^{2/3}\,
\left(\frac{f}{1\,{\rm kHz}}\right)^{-1/2}\,{\rm sec}
& (m_2 \simeq m_{\rm NS})\nonumber\\
4.4\times 10^2\,\left(\frac{P}{10\,{\rm days}}\right)^{2/3}\,
\left(\frac{m_{2}}{10\,\Ms}\right)^{-1/6}\,
\left(\frac{f}{1\,{\rm kHz}}\right)^{-1/2}\,{\rm sec}
& (m_2 \gg m_{\rm NS})\\
\end{array}
\right.
\,;
\label{Tmax1}
\ee
For comparison, we note that the maximum integration time before
which the motion of the detector from Earth's rotation about its axis,
produces a Doppler shift in frequency greater than the frequency resolution 
bin is $T_{\rm max}^{\rm (det)} \simeq 50\, (f/1\,{\rm kHz})^{-1/2}$ min.  
It is then evident that in the case of a CW source orbiting a companion object, 
the frequency shift is typically more severe and the computational challenge 
that one faces when searching for the additional 5-dimensional parameter 
vector describing the binary orbit is considerable.

\subsection{Expected sources and motivation}
\label{subs:sm}

The {\it characteristic amplitude} of the continuous signal emitted by a 
triaxial NS, rotating about a principal axis with period $P_r$ and 
frequency $f_r = 1/P_r$ reads~\cite{Thorne87}:
\be
h_{\rm c} \simeq 7.7 \times 10^{-26}\,
\left(\frac{\epsilon_e}{10^{-6}}\right)\,
\left(\frac{D}{10\,{\rm kpc}}\right)^{-1}\,
\left(\frac{f_r}{500\,{\rm Hz}}\right)^{2} ,
\label{hchar}
\ee
where $\epsilon_e$ is the equatorial ellipticity of the star, and we
have taken the standard value $10^{45}$ gr cm$^2$ for the NS moment of
inertia. In Eq.~(\ref{hchar}) we assume that the GW energy is released at
exactly twice the rotation frequency, so that $f = 2 f_r$. 

\begin{figure}
\begin{center}
\epsfig{file=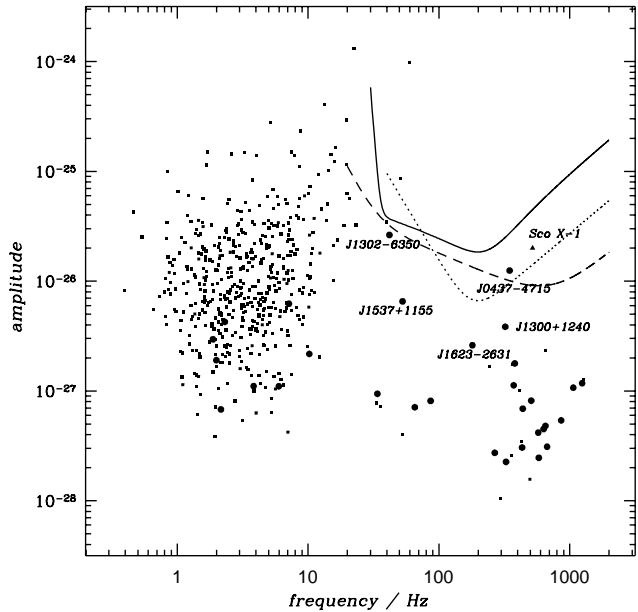,width=4.1in}
\caption{\label{fig:h}
Upper limits on the GW amplitude from known radio pulsars and Sco X-1.
The plot shows the upper-limit on $h_{\rm c}$, Eq.~(\protect{\ref{hchar1}}),
for the neutron stars included in the latest version of the published 
radio pulsar catalogue~{\protect\cite{cat}},
for which the spin-down and the distance are known (isolated pulsars: squares;
binary pulsars: bold bullets), and the estimated value of $h_{\rm c}$ for
Sco X-1 (triangle), Eq.~(\protect{\ref{hchar2}}). We also show the noise amplitude 
$h_{\rm n}$ for an observation
time $T = 10^7$ sec, for the first generation detectors (GEO600: solid line;
LIGO I: dotted line; VIRGO: dashed line).}
\end{center}
\end{figure}

The characteristic amplitude has to be compared with the noise
fluctuations, 
\be
h_{\rm n} = \sqrt{\frac{S_n(f)}{T}} = 10^{-25}\,
\left(\frac{S_n(f)}{10^{-43}\,{\rm Hz}^{-1}}\right)^{1/2}\,
\left(\frac{T}{10^7\,{\rm sec}}\right)^{-1/2}\,
\label{hn}
\ee
in a resolution bin, where $S_n(f)$ is the noise spectral density of
the instrument. A signal is detected if $h_{\rm c} \simgt k h_{\rm n}$,
where the multiplicative factor $k$, of order of a few, 
depends on the number of filters used in the analysis and on the
(subjective) confidence level of the measurement. The strength of the signal
depends crucially on the $\epsilon_e$, which is a measure
of the non-axis-symmetry of a star. Although large
uncertainties exist in the maximum value that $\epsilon_e$ can achieve,
theoretical studies of the braking strain of NS crusts suggest that
$\epsilon_e \simlt 10^{-5}$. One can observationally constrain 
$\epsilon_e$ if the NS rotational period $P_r$ and its first time derivative
$\dot{P_r}$ are known:
\be
\epsilon_e \le 5.1 \times 10^{-7}\,
\left(\frac{P_r}{2\,{\rm msec}}\right)^{3/2}\,
\left(\frac{\dot{P_r}}{10^{-15}}\right)^{1/2}\, .
\label{ell}
\ee
This yields an upper-limit on the amplitude of the emitted GW's, by assuming
that all energy is lost through gravitational wave emission. By combining
Eqs.~(\ref{hchar}) and~(\ref{ell}), we can immediately place the upper-limit
\be
h_{\rm c} \simlt 3.9\times 10^{-26}\,
\left(\frac{f_r}{500\,{\rm Hz}}\right)^{1/2}\,
\left(\frac{\dot{P_r}}{10^{-15}}\right)^{1/2}\,
\left(\frac{D}{10\,{\rm kpc}}\right)^{-1}\, .
\label{hchar1}
\ee
The comparison of $h_{\rm c}$ for some known NS's with $h_{\rm n}$ of the
laser interferometers is shown in Fig.~\ref{fig:h}.

In Table~\ref{tab:pulsars} we also show some of the parameters
of the binary radio pulsars which are known so far. It is interesting to 
notice that a bunch of systems are within the
reach of the first generation of instruments for coherent observation
times from 15 months to 4 years, and well within the enhanced configuration
sensitivity. It is also worth noticing that they represent a fair sample
of orbital companions, as they include NS/NS binaries, NS/white-dwarf
binaries, NS/main-sequence stars and NS/planet. However, the actual 
signal strain could be orders of magnitude below
the upper-limit~(\ref{hchar1}).

A far more promising scenario is the accretion of hot material onto the 
NS surface. Here the induced quadrupole moment is directly
related to the accretion rate, which can be copious. The gravitational 
energy reservoir, moreover, can be continuously
replenished, if persistent accretion occurs. The key idea behind
this scenario is that
gravitational wave radiation can balance the torque due to accretion, and
was proposed over 20 years ago~\cite{PP78,Wagoner84}. However, it
has attracted considerable new interest in the past two years and has been
fully revitalized  by the launch of the Rossi X-ray Timing Explorer, designed 
for precision timing of accreting NS's. %
\vbox{
\begin{table}
\begin{center}
\caption{\label{tab:pulsars}
Neutron stars detected as radio pulsars in binary systems. The table shows
the main parameters of the 44 known neutron stars detected as radio
pulsars orbiting a companion object listed in the Taylor et 
al. catalogue~{\protect\cite{cat}}. We list the gravitational wave frequency
$f$ -- assumed to be exactly twice the radio frequency -- its first
time derivative $\dot{f}$, the estimated distance
$D$ from the solar system and the spin-down age (we do not give all 
significant digits). The orbital parameters -- the orbital period
$P$, the projection of the semi-major axis along the line of sight $a_{\rm p}$
and the eccentricity $e$ -- are tabulated with their respective errors 
(in parenthesis). Notice that out of the 44 known systems,
seven sources radiate at a frequency $< 10$ Hz and
therefore are not observable by the (so far) planned ground-based
experiments.
}
\begin{tabular}{ll|cccc|ccc}
PSR J & PSR B & $f$      & $\dot{f}$              & $D$ & $\log({\rm age}/{\rm yr})$ & $P$  & $a_{\rm p}$ & $e$ \\
      &       & ${\rm Hz}$ & $10^{-15}\,{\rm sec}^{-2}$ & kpc &                       & days & sec              &     \\
\hline
 0024-7204E   & 0021-72E &  565.6   &    --    &  4.5  &  --     &    2.256844 (3)       &     1.978 (3)      &   0.000 (3) \cr
 0024-7204I   & 0021-72I &  573.9   &    --    &  4.5  &  --     &    0.226 (3)          &     0.039 (2)      &   0.00 (5) \cr
 0024-7204J   & 0021-72J &  952.1   &    --    &  4.5  &  --     &    0.120665 (2)       &     0.0402 (2)     &   0.00 (5) \cr
 0034-0534    &          &  1065.4  &     3.8  &  1.0  &    9.6  &    1.58928180 (3)     &     1.437768 (5)   &   0.0000 (1) \cr
 0045-7319    &          &  2.2     &    10.5  & 57.0  &    6.5  &   51.169226 (3)       &   174.2540 (8)     &   0.807995 (5) \\ 
 0218+4232    &          &  860.9   &    27.8  &  5.8  &    8.7  &    2.02885 (1)        &     1.9844 (1)     &   0.00000 (2) \\
 0437-4715    &          &  347.4   &     3.4  &  0.1  &    9.2  &    5.741042329 (12)   &     3.3666787 (14) &   0.0000187 (10) \cr
 0613-0200    &          &  653.2   &     2.3  &  2.2  &    9.6  &    1.19851251 (2)     &     1.09145 (2)    &   0.000000 (22) \\
 0700+6418    & 0655+64  &  10.2    &     0.04 &  0.5  &    9.7  &    1.028669703 (1)    &     4.125612 (5)   &   0.0000075 (11) \cr
 0751+1807    &          &  574.9   &     1.3  &  2.0  &    9.8  &    0.2631442679 (4)   &     0.396615 (2)   &   0.0000 (1) \\
 0823+0159    & 0820+02  &  2.3     &     0.3  &  1.4  &    8.1  & 1232.47 (12)          &   162.1466 (7)     &   0.011868 (7) \cr
 1012+5307    &          &  380.5   &     1.1  &  0.5  &    9.8  &    0.604672713 (5)    &     0.581816 (6)   &   0.00000 (2) \\
 1022+10      &          &  121.6   &    --    &  0.6  &  --     &    7.80513014 (8)     &    16.765412 (6)   &   0.0000977 (6) \cr
 1045-4509    &          &  267.6   &     0.7  &  3.2  &    9.8  &    4.0835291 (1)      &     3.015107 (9)   &   0.000019 (6) \cr
 1300+1240    & 1257+12  &  321.6   &     5.9  &  0.6  &    8.9  &   66.536 (1)          &     0.0013106 (6)  &   0.0182 (9) \cr
 1302-6350    & 1259-63  &  41.9    &  1994.5  &  4.6  &    5.5  & 1236.72359 (5)        &  1296.580 (2)      &   0.869931 (1)   \cr
 1312+1810    & 1310+18  &  60.3    &    --    & 18.9  &  --     &  255.8 (6)            &    84.2 (7)        &   0.002 (3) \cr
 1455-3330    &          &  250.4   &    --    &  0.7  &  --     &   76.17458 (2)        &    32.36222 (6)    &   0.000167 (2) \cr
 1518+4904    &          &  48.9    &    --    &  0.7  &  --     &    8.6340047 (8)      &    20.043982 (8)   &   0.2494848 (7)    \cr
 1537+1155    & 1534+12  &  52.8    &     3.4  &  0.7  &    8.4  &    0.4207372998 (3)   &     3.729468 (9)   &   0.2736779 (6)   \cr
 1623-2631    & 1620-26  &  180.6   &    12.9  &  1.8  &    8.3  &  191.442819 (2)       &    64.809476 (13)  &   0.0253147 (5) \cr
 1640+2224    &          &  632.2   &     0.6  &  1.2  &   10.2  &  175.460668 (2)       &    55.329722 (2)   &   0.00079742 (5) \cr
 1641+3627B   & 1639+36B &  566.9   &    --    &  7.7  &  --     &    1.259113 (3)       &     1.389 (2)      &   0.005 (5) \cr
 1643-1224    &          &  432.7   &     3.1  &  4.9  &    9.3  &  147.01743 (6)        &    25.07260 (4)    &   0.000506 (2) \cr
 1713+0747    &          &  437.6   &     0.8  &  1.1  &    9.9  &   67.82512988 (2)     &    32.342413 (1)   &   0.00007492 (2) \cr
 1721-1936    & 1718-19  &  2.0     &     3.2  & 11.6  &    7.0  &    0.2582735 (2)      &     0.3526 (8)     &   0.000 (5) \\
 1748-2446A   & 1744-24A &  173.0   &    -0.3  &  7.1  &  --     &    0.0756461170 (4)   &     0.11963 (2)    &   0.0000 (12) \\
 1803-2712    & 1800-27  &    6.0   &     0.3  &  3.6  &    8.5  &  406.781 (2)          &    58.9397 (18)    &   0.000507 (6) \cr
 1804-0735    & 1802-07  &  86.6    &     1.8  &  3.2  &    8.9  &    2.6167634 (5)      &     3.92047 (4)    &   0.211999 (15)   \cr
 1804-2718    &          &  214.1   &    --    &  1.2  &  --     &   11.15 (1)           &     7.20 (4)       &   0.00 (1) \\
 1823-1115    & 1820-11  &  7.1     &    35.2  &  6.3  &    6.5  &  357.7622 (3)         &   200.672 (3)      &   0.79462 (1)    \cr
 1834-0010    & 1831-00  &  3.8     &     0.1  &  2.6  &    8.8  &    1.811103 (2)       &     0.7231 (8)     &   0.000 (4) \\
 1857+0943    & 1855+09  &  373.0   &     1.2  &  0.9  &    9.7  &   12.3271711905 (6)   &     9.2307802 (4)  &   0.00002168 (5) \cr
 1910+0004    &          &  552.7   &    --    &  4.1  &  --     &    0.140996 (1)       &     0.038 (2)      &   0.00 (1) \\
 1915+1606    & 1913+16  &  33.9    &     4.9  &  7.1  &    8.0  &    0.322997462736 (7) &     2.3417592 (19) &   0.6171308 (4)    \cr
 1955+2908    & 1953+29  &  326.1   &     1.6  &  5.4  &    9.5  &  117.349097 (3)       &    31.412686 (5)   &   0.0003304 (3) \cr
 1959+2048    & 1957+20  &  1244.2  &    13.0  &  1.5  &    9.2  &    0.3819666389 (13)  &     0.892268 (8)   &   0.00000 (4) \\
 2019+2425    &          &  508.3   &     0.9  &  0.9  &    9.9  &   76.5116347 (1)      &    38.767625 (1)   &   0.00011111 (6)\cr
 2033+17      &          &  336.2   &    --    &  1.4  &  --     &   56.2 (1)            &    20.07 (8)       &   0.00 (5) \\
 2130+1210C   & 2127+11C &  65.5    &    10.7  & 10.0  &    8.0  &    0.335282052 (6)    &     2.520 (3)      &   0.68141 (2)    \cr
 2145-0750    &          &  124.6   &    --    &  0.5  &  --     &    6.83890256 (8)     &    10.16411 (1)    &   0.000021 (2) \\
 2229+2643    &          &  671.6   &     0.4  &  1.4  &   10.4  &   93.015895 (2)       &    18.912519 (5)   &   0.0002550 (4) \cr
 2305+4707    & 2303+46  &  1.9     &     1.0  &  4.4  &    7.5  &   12.33954454 (17)    &    32.6878 (3)     &   0.658369 (9)    \cr
 2317+1439    &          &  580.5   &     0.4  &  1.9  &   10.4  &    2.459331464 (2)    &     2.3139483 (9)  &   0.0000005 (7) \cr
\end{tabular}
\end{center}
\end{table}%
}
The observational evidence that Low
Mass X-ray Binaries (LMXB's) - binary systems where a compact object accretes material
from a low mass companion -- in our Galaxy are clustered around a rotation frequency
$\approx 300$ Hz, led Bildsten~\cite{Bildsten98} to propose a mechanism to
explain this behaviour. 
The fundamental idea is that continuous emission of GW's
radiates away the angular momentum that is transfered to the NS by the infalling
material. The fact
that the rate of angular momentum loss through GW's scales as $f^5$, provides
a very natural justification of the clustering of rotation frequency of several
sources. The physical process responsible for producing a net quadrupole moment
is the change of composition in the NS crust, which in turn is produced
by the temperature gradient caused by the in-falling hot material.  
Recently, Ushomirsky et al.~\cite{UCB00} 
have posed this initial idea on more solid theoretical grounds. If such mechanism
does operate, LMXB's are extremely interesting candidate sources for 
Earth-based detectors. Several systems would be detectable by LIGO 
operating in the "enhanced" configuration (LIGO II), if the detector sensitivity
is tuned, through narrow-banding, around the emission frequency. In particular,
Sco X-1, the most luminous X-ray source in the sky, possibly is marginally 
detectable by "initial" LIGO and GEO600 (the latter in narrow-band 
configuration), where  
an integration time of approximately 2 years would be required. The 
characteristic amplitude for this class of sources is:
\be
h_{\rm c} \simeq 4\times 10^{-27}\,\left(\frac{R_{NS}}{10^6\,{\rm cm}}\right)^{3/4}\,
\left(\frac{m_{NS}}{1.4\,\Ms}\right)^{-1/4}\,
\left(\frac{F_X}{10^{-8}\,{\rm erg}\,{\rm cm}^{-2}\,{\rm sec}^{-1}}\right)^{-1/2} ,
\label{hchar2}
\ee
where $R_{NS}$ is the typical radius of a neutron star and $F_X$ is the X-ray flux.

Data analysis strategies to search for NS's in binary systems might therefore be
vital for ground-based detectors. They would supplement the already existing 
strategies for the several sources mentioned above.

\subsection{Organization of the paper}

The paper is organized as follows: 
In Section \ref{sec:model} we set up the parametrized model of the signal based 
on the two body problem. We then identify the search parameters of the signal, 
which are essentially the binary orbital elements, for the general 
case of elliptic orbits, as well as for circular orbits. In Section 
\ref{sec:datan} we discuss the geometrical approach to the data analysis;
in Section~\ref{subs:nfilt} we review 
the method of determining the number of templates which involves the computation
of the metric on the signal manifold, and we also state useful 
properties of the metric which are then used in the later sections. In 
Section~\ref{subs:tar} we present a rigorous criterion, naturally suggested 
by the geometrical picture, for estimating the number of filters required in 
targeted searches and to determine whether any particular source is a 
one-filter target.

Sections~\ref{sec:cost} and~\ref{sec:obs} contain the key results of the paper. 
In Section~\ref{sec:cost}
we obtain analytic approximate formulae for the number density of 
filters per unit coordinate volume in the parameter space in the two important 
regimes of observation time: (i) observations lasting for several orbital 
periods, and (ii) observations lasting a fraction of an orbit. These 
expressions provide for us the scaling: that is, how the computational cost 
scales with different parameters. They can also be readily applied to estimate 
the computational cost of detecting GW from some observed systems. Further, 
we also compare these results with the fully numerical computation of the 
metric determinant (which shows excellent agreement over a wide range of the 
parameters), and determine the range of applicability of these approximate 
expressions as a function of the relevant parameters (in particular the time 
of observation as compared to the orbital period). In the Section~\ref{sec:obs} 
we apply the above analysis to known NS's in binary systems, specifically 
radio-pulsars and LMXB's. Given the current (or near future) computational 
resources, we then determine the systems that could be searched by 
matched filtering methods. Section~\ref{sec:concl} contains our final
remarks and future directions of work.

%
%

\section{The model}
\label{sec:model}

\subsection{The description of the problem}
\label{subs:description}

In this paper we employ the matched filtering technique for coherent integration 
of the signal. It provides for us the optimal signal-to-noise ratio that can be
achieved with linear data analysis techniques, and also an elegant theoretical 
framework can be given in which the computational costs can be derived in a very 
transparent way. We have already remarked that the processing power of 
present/near future computers falls woefully short for carrying out full 
matched-filter based searches of CW sources. However, it is important to note 
that matched filters are implemented in the final stage of every hierarchical 
strategy that has been devised so far~\cite{BC00,SP99}.
Our analysis can be therefore applied to the typically reduced parameter 
range which is relevant in this final step.
Since the final stage accounts for most of the processing power required in the entire 
search, our results essentially reflect the computational costs involved in the 
search. 

The amplitude of the GW's is expected to be small for CW sources. Therefore long 
observation times are imminent for extracting the signal out of the detector noise 
with a reasonable signal-to-noise ratio (SNR). The 
observation times could last for a few months, up to a year or more. During 
this period of observation it becomes essential to take into account the 
changes in frequency which result from the variation in relative positions of 
the source and the detector -- Doppler effects -- and also those that are 
intrinsic to the source, such as the spin-down. The general idea is 
(i) to correct for these changes in the phase, 
(ii) use the FFT algorithm to compute efficiently the power spectrum and then, 
(iii) look for a statistically significant peak in the power spectrum
itself.

The aim of this paper is to estimate the extra computational costs 
associated with the search for sources in binary systems. In order to disentangle
the phase Doppler modulation produced by the source motion
around the orbital companion, and limit the complexity of our analysis, we will
make the following assumptions:
\begin{itemize}

\item The GW source radiates perfectly monochromatic GW's at the intrinsic 
frequency $f_0$, that is the 
frequency in its own rest frame is constant: we ignore the effect of
spin-down parameters.

\item The location of the source in the sky is perfectly known: the Doppler
effect produced by the motion of the Earth spinning on its own axis, orbiting 
the Sun etc. can be computed and subtracted from the data stream
with very little CPU load.

\end{itemize}
These hypotheses leave us only to grapple with the problem of the motion of the 
source. Notice that they do not affect the generality of our analysis. This 
approach returns the extra cost involved in searching for NS's in binary 
systems, for each template of the isolated NS search (with the appropriate number
of spin-down parameters and range of parameter values). Our analysis thus provides
a solid upper-limit on the total computational costs (it ignores possible 
correlations between the parameters). In fact, if we call ${\cal N}_{\rm is}$ 
the number of templates to search for an isolated NS (the search parameters
are the location in the sky and spin-down coefficients), and ${\cal N}$ the number
of templates required to search for a 
perfectly monochromatic source in binary orbit whose location in the sky is 
assumed to be known, the search parameters being only the orbital elements 
of the binary system, then the total number of filters
${\cal N}_{\rm T}$ required for a generic source, is bound by the 
following inequality:
\be
{\cal N}_{\rm T} \leq {\cal N}_{\rm is} \times {\cal N}\,.
\label{N_is+b}
\ee

We take the orbit of the source to 
be Keplerian, elliptical in shape, but we assume that we do not exactly 
know, or do not know at all, the orbital parameters. It is possible that a 
particular parameter or a set of parameters is known so accurately that no 
search is required over these parameters. Such a situation arises for some of 
the known radio binary pulsars listed in the published catalogue~\cite{cat}. 
We then have a {\it targeted search}, where one 
must now search only over rest of the parameters. In the extreme case, all 
parameters may be known so accurately, that only a single Doppler correction 
needs to be 
applied (one filter in the language of data analysis). Usually for known systems, 
some orbital elements are known within given error bars. For example, 
one might know the period of the orbit within certain limits, then these limits 
provide for us the ranges of the parameters, over which 
the search must be carried out. On the other hand we might not know anything 
at all about the binary system, in which case one must launch a more extensive 
search with a large number of templates covering the entire region of the 
parameter space - this is the so-called {\it ``blind'' search}.

\subsection{The two body problem, orbital elements and search parameters}
\label{subs:orbpar}

The essential problem of celestial mechanics is the two-body problem, where 
one solves for the motion of two point particles of masses $m_1$ and $m_2$,
and position vectors $\r_1$ and $\r_2$, attracted by their mutual gravitational 
force. The solution is simple in their centre-of-mass frame in terms of the 
relative position vector $\r = \r_1 - \r_2$. The vector $\r$ lies in a 
plane and traces out an ellipse; the individual masses trace out 
similar ellipses scaled by factors depending on the masses:
\be
\r_1 = {m_2 \over M}\, \r\,,\quad\quad\r_2 = - {m_1 \over M}\, \r\,,
\label{r12}
\ee
where $M = m_1 + m_2$. Here, we (arbitrarily) take $\r_1 \equiv \r_{\rm NS}$ 
to describe the position of the 
CW source we want to detect. We assume that the centre-of-mass of the 
binary system can be considered at rest, or in uniform motion, with respect
to the Solar System barycentre during the period of observation; therefore 
knowing $\r_1$ as a function of time is sufficient for us to compute the 
GW Doppler phase shift $\phi_D(t)$.

It is convenient to set up a Cartesian coordinate system 
$(\xi, \eta, \zeta)$ attached to the binary source, so that: 
(i) $\r_1$ lies in the $(\xi, \eta)$ plane with the origin at the 
centre of the ellipse; 
(ii) the semi-major axis of the ellipse coincides with the $\xi$-axis, and 
(iii) the $\zeta$-axis is perpendicular to the orbital plane and points in the 
direction of the orbital angular momentum. We specify the direction to the 
detector, in the $(\xi, \eta, \zeta)$ frame, by the unit vector:
\be
\n = (\sin \epsilon \cos \psi)\,\gras{\hat{\xi}} + 
(\sin \epsilon \sin \psi) \,\gras{\hat{\eta}}
+ (\cos \epsilon) \,\gras{\hat{\zeta}}\,,
\label{n}
\ee
where $\epsilon$ and $\psi$ are the usual polar angles.

The Doppler phase shift at the detector output therefore reads:
\be
\phid (t) = - \frac{2 \pi f_0}{c}\,\left[ \n \cdot \r_1 (t)\right]\,,
\label{doppler}
\ee
where $c\simeq 2.9979 \times 10^{10}$ cm/sec is the speed of light. 

The orbit in the $(\xi, \eta)$ plane is given as follows:
Let $a$ be the {\it semi-major axes} of the elliptical orbit and $e$ the 
{\it eccentricity}, then, the orbit is described by the equations:
\bea
\xi(t) &=&  a \cos E(t), \nonumber\\ 
\eta(t) &=& a \sqrt{1 - e^2}\, \sin E(t), 
\label{orbpl}
\eea
where $E$ is the so-called {\it eccentric anomaly}, and is a function of the 
time $t$. It is related to the {\it mean angular velocity} $\omega$ and the 
{\it mean anomaly} $M(t)$ by the Kepler equation,
\be
\label{kepler}
E(t) - e \sin E(t) = \omega t + \a \equiv M(t),
\ee  
where $\a$ is an initial phase, $0 \le \a < 2 \pi$. When $\omega t + \a = 0$ we 
have $E = 0$ and the mass is closest to the focus $\xi = a e, \eta = 0$. These 
equations describe the orbit in the $(\xi, \eta)$-plane as function of time,
that is determined by the four orbital elements 
$a, \omega, \a, e$. However, the orbit in space requires two additional 
parameters, namely, the angles $\epsilon$ and $\psi$ mentioned before.
Thus, in all we have {\it six orbital elements which specify the orbit in space}.

The Doppler phase correction, Eq.~(\ref{doppler}), is obtained 
from Eq.~(\ref{orbpl}), and reads as follows:
\be
\phid(t) = - {2 \pi f_0 a \sin \epsilon \over c}\,\left[\cos \psi \cos E(t) + \sin \psi 
\sqrt{1 - e^2} \sin E(t)\right]. 
\label{doppler1}
\ee 
The total phase $\Phi$ at the barycentre (we have assumed that the 
corrections have been made for the Earth's motion) is therefore,
\be
\Phi(t) = 2\pi f_0 t + \phid (t)\, .
\label{phiT}
\ee
Since we have assumed that we have corrected for the Earth's motion, the $t$ 
can be regarded as the barycentric time.  

For computing the metric in the next section we will require the time averages of 
the derivatives of the phase with respect to the search parameters. Since the 
Kepler equation connects $E$ to the time $t$, or $M$ implicitly, $\cos E$ and 
$\sin E$ are usually expressed as power series in the eccentricity parameter 
$e$ with harmonics in $M$. Also the time averages are more conveniently computed 
using $M$. We thus write formally,
\bea
\cos E &=& \sum_{k = 0}^{\infty} \Ck(e) \cos(k M)\,, \nonumber\\
\sqrt{1 - e^2} \sin E &=& \sum_{k = 1}^{\infty} \Sk(e) \sin(k M)\,, 
\label{pwrser}
\eea
where $\Ck (e)$ and $\Sk (e)$ are power series in $e$. In this work we consider  
expansions up to the 7th power in $e$, which automatically means that we 
consider 7 harmonics in $M$. The $\Ck(e)$ and $\Sk(e)$ are given in 
Appendix~\ref{app:CS} upto this order.
 
The parameters over which one must launch a search are not exactly the 
orbital elements, and need not be of the same number. It is the Doppler phase 
correction that is observed and so the information about the system that we 
can glean depends on the combination of the orbital elements that enter into it. 
The $a$ and $\epsilon$ combine into a single parameter 
$a \sin \epsilon \equiv a_{\rm p}$, the projected semi-major axis along the line 
of sight, which is actually the quantity inferred from astronomical observations.
The other search parameters are the remaining orbital elements 
$\omega, \a, e$ and $\psi$. So {\it in the general case}, when we do not know 
any of the parameters exactly, we have a {\it 5-dimensional parameter space} to 
search. 

We note that $f_0$ is not a search parameter because of the special search 
technique that is employed~\cite{BCCS98}: it involves the
"stretching" of the time coordinate in such a way, so as to make the signal 
appear monochromatic in this time coordinate. One then simply takes the FFT to 
compute the power spectrum, which now is concentrated in a single frequency bin.

The gravitational waveform that we are dealing with, reads (in the 
barycentric frame):
\be
h(t, \L) = \Re [{\cal A} \exp (- i \Phi (t; \L) + i \Psi )],
\label{hFFT}
\ee
where $\Phi (t; \L)$ is given by Eq.~(\ref{phiT}), 
$\l = (a_p, \omega, \a, e, \psi)$ is the 5-dimensional vector that refers to
the parameters that are required in the discrete mesh search, 
and $\L = (f_0, \l)$. 
The polarization amplitude ${\cal A}$, and the polarization phase $\Psi$ are 
slowly varying time-dependent functions over the time-scale of the day 
and depend on the relative orientations of the source and the detector. In
agreement with all the investigations carried out so far, we assume them
constant in our analysis. It is expected that 
these factors can be easily included in the full analysis and will not  
significantly affect the computational burden~\cite{BCCS98}.

%
%

\section{The data analysis}
\label{sec:datan}

In the geometrical picture \cite{DS94,BSD96}, the signal is a vector in 
the vector space of data trains and the $N$-parameter family of signals traces out
an $N$-dimensional manifold which is termed as the {\it signal manifold}. The 
parameters themselves are coordinates on this manifold. One can introduce a 
metric $\gamma_{jk}$ on the signal manifold which is related to the fractional 
loss in the SNR when there is a mismatch of parameters between the 
signal and the filter. The spacing of the grid of filters is decided by the 
fractional loss due to the imperfect match that can be tolerated. 
Given the parameter space that one needs to scan, it is then easy to
estimate the total number of filters required to carry out the search for the 
signal. In Sec.~\ref{subs:nfilt} we first briefly 
review the method introduced by Brady et al.~\cite{BCCS98} which in turn was 
based on Owen's~\cite{Owen96} method for searching for GW signals from 
in-spiralling compact binaries. We also present some useful general properties 
of the metric on the signal manifold which are used in the next section.
In Sec.~\ref{subs:tar} we then consider
the case where accurate information about one or more parameters of a NS
are available (for instance through
radio observations) so that the parameter space that one needs to search over, is
drastically reduced. We present a rigorous approach, based on  
differential geometric methods, to determine the exact dimensionality of the 
data analysis problem and formulate criteria to decide whether a given source is
a one-filter target.

\subsection{Number of filters: General}
\label{subs:nfilt}

In the method presented in~\cite{BCCS98}, the key idea is to first correct for 
the Doppler effect in the phase of the signal for each of the grid points of the 
parameter space and then compute the power spectrum. The latter is obtained 
efficiently via the FFT algorithm. Even if we know the frequency of the pulsar, 
it is desirable to search over a band $\simeq 1\%$ of the pulsar 
frequency~\cite{JS}. We therefore have a large number of frequency bins to 
search over and the FFT algorithm is thus computationally advantageous.   
If the Doppler correction is right, that is, if the signal and 
filter parameters match perfectly, then the signal is all concentrated at 
$f = f_0$ in the power spectrum. The grid spacing is decided by the amount the 
maximum of the power spectrum falls, when the parameters of the signal and 
filter mismatch. The mismatch $\mu$ is defined as the fractional reduction in 
the maximum of the power spectrum when the parameters mismatch. Fixing the 
mismatch $\mu$, fixes the grid spacing of the filters in the parameter space 
which we will denote by $\cal P$. The number density of filters 
(the number of filters per unit proper volume - proper volume defined through 
the metric) in $\cal P$ depends on $\mu$, and is denoted by $\rho_N (\mu)$, where
$N$ is the dimension of $\cal P$. For a hyper-rectangular mesh (which does not 
represent necessarily the most efficient tiling of the parameter space) it reads:
\be
\rho_N (\mu) = \left [ \h \sqrt {N \over \mu} \right ]^{N}\,.
\label{rho}
\ee
In fact, the {\it proper distance} $dl$ between two filters is:
\be
dl = 2\sqrt{\frac{\mu}{N}}\simeq 
0.15 \sqrt{\left(\frac{\mu}{3\%}\right)\,\left(\frac{5}{N}\right)}\,.
\label{dl}
\ee
The proper volume of $\cal P$, can be easily computed from $\gamma_{jk}$~\cite{BCCS98}:
\be
\Vp = \int_{{\cal P}}d \l\, \sqrt{\det ||\gamma_{jk}||} \,;
\label{vp}
\ee
the number of filters ${\cal N}$ is then just the proper volume~(\ref{vp}), times 
the filter density~(\ref{rho}):
\be
{\cal N} = \rho_N (\mu) \Vp\,.
\label{Nf}
\ee
For the elliptical orbit we have $N = 5$, while for the circular case the number 
reduces to $N = 3$. See Table~\ref{tab:rho} for the relevant values of 
$\rho_N (\mu)$.

\vbox{
\begin{table}
\begin{center}
\caption{\label{tab:rho}
The number density of filters $\rho_N(\mu)$, Eq.~(\protect{\ref{rho}}) for the circular and
elliptical orbit case (the dimensions of the parameter space are 
$N=3$ and $N=5$, respectively) as a function of the {\it mismatch} $\mu$.
}
\begin{tabular}{c|ccccc}
dimension of the     &   \multicolumn{5}{c}{mismatch} \\
\cline{2-6}
parameter space      & $1\%$    &    $3\%$  &    $5\%$  &  $10\%$ &    $30\%$ \\
\hline
 3                   & 649.5    &   125.0    &    58.1  &   20.5  &     3.9  \\
 5                   & 174963.0 & 11206.5    &  3125.0  &  552.4  &    35.4  \\
\end{tabular}
\end{center}
\end{table}%

If the signal parameters are $\L = (f_0, \l)$ and the filter parameters are
$\l + \bdl$, the power spectrum for an observation time $T$ is given by,
\be
P(f;f_0, \l, \bdl) = {{\cal A}^2 \over T} 
\left \vert \int_0^T dt\,\exp [i \Phi (t; \l, \bdl)]  \right \vert^2\,,
\label{pwspc}
\ee
where
\be
\Phi (t; \l, \bdl) = 2 \pi (f - f_0) t + \phid (t; f_0, \l + \bdl) - 
\phid (t; f_0, \l)\,.
\label{phase}
\ee
The mismatch is both in $\l$ as well as in $f$ (this can occur because of sampling 
at the wrong frequency) and is denoted by $m (\L, \bdL)$:
\be
\label{mm}
m(\L, \bdL) \equiv
1 - {P (f; f_0, \l, \bdl) \over P (f_0; f_0, \l, \bar {0})}
 \simeq g_{\a \b} ( \L) \Delta \Lambda^{\a} \Delta \Lambda^{\b} + o (\Delta \L ^3)\,.
\ee
From Eqs.~(\ref{pwspc}) and~(\ref{mm}) the metric $g_{\a \b}$ can be computed 
by Taylor expansion. It is given by,
\be
\label{metric}
g_{\a \b} = \m \Phi_{\a} \Phi_{\b} \M - \m \Phi_{\a} \M  \m\Phi_{\b} \M , 
\ee
where the suffix, say $\a$, denotes derivative with respect to $\dL^{\a}$ 
and the angular brackets denote time averages defined as follows: For a function 
$X (t)$ defined on the data train $[0, T]$, the time average of $X$ is, 
\be
\m X \M = {1 \over T} \int_0^T\,dt\, X(t)\,.
\label{Xav}
\ee
We remark that $g_{\a \b}$ is not the metric which is used to calculate the 
proper volume, because it still includes $f$. We need 
to maximize over $f$, which is tantamount to projecting $g_{\a \b}$ orthogonal 
to the $\Delta f$ direction. Thus the metric on the sub-manifold of the search 
parameters $\l$ is, 
\be
\gamma_{jk} = g_{jk} - {g_{0j} g_{0k} \over g_{00}}\,.
\label{gamma}
\ee 
Here the Greek and Latin indices range over the parameter
$\L$ ($\a,\b = 0,1,...,N$) and $\l$ ($j,k = 1,2,...,N$), respectively;
the index $0$ identifies the parameter corresponding to the frequency $f$.
If $\fM$ is the highest GW frequency that we are searching for, then 
we must put $f_0 = \fM$ in the above expression for $\gamma_{jk}$. The 
proper volume of the signal and the total number of templates can be 
easily derived by inserting Eqs.~(\ref{rho}) and~(\ref{gamma}), 
into Eqs.~(\ref{vp}) and~(\ref{Nf}), respectively. Notice that the parameter
$\mu$ is the projected mismatch, where the maximization over $f$ has already been 
accounted for. 
}

The metric defined above through the phase $\Phi$ has certain elegant 
properties which we will use in the next section to simplify computations:
\begin{enumerate}

\item {\em{Scaling}}: If we scale $\Phi$ by a constant factor $\chi$, that is if 
we define, 
\be
\Phi = \chi \tPhi\,,
\label{sc_phi}
\ee
then each component of the metric $g_{\a \b}$ and 
$\gamma_{jk}$ is scaled by the factor $\chi^2$:
\be
g_{\a \b} = \chi^2\, \tilde g_{\a \b} \,,
\quad\quad
\gamma_{jk} = \chi^2\, \tilde \gamma_{jk} \,;
\label{sc_g}
\ee
the determinants are scaled accordingly:
\be
\det ||g_{\a \b}|| = \chi^{2N+2}\, \det ||\tilde g_{\a \b}|| \,,
\quad\quad
\det ||\gamma_{jk}|| = \chi^{2N}\, \det ||\tilde \gamma_{jk}|| \,;
\label{sc_d}
\ee
the proper volume $\Vp$ is scaled by the square root 
of the determinant, and so is ${\cal N}$:
\be
\Vp = (\chi^N)\, \tilde \Vp\,,
\quad\quad
{\cal N} =  (\chi^N)\, \tilde {\cal N}\,.
\label{sc_v}
\ee

\item {\em{Translation}}: If a function only of the parameters, say $f(\L)$, 
{\em not containing time} is added to $\Phi$, the metric remains invariant under this 
transformation. 

\end{enumerate}
These properties are easily verified.
We use the first property to make all the coordinates dimensionless in the 
metric, compute $\tVp$,  the proper volume in the scaled metric, and then 
finally multiply by the appropriate power of $\chi$ to obtain the actual volume 
$\Vp$ and also the number of filters. We have already used the second property 
to shift the origin  from the focus to the centre of the elliptical orbit in 
Eq. (\ref{orbpl}). The expressions for the phases differ by an amount 
proportional to $a e$, which leaves the metric unaltered. 

\subsection{Targeted searches}
\label{subs:tar}

So far we have assumed that the size of the parameter space
${\cal P}$ along any direction is much larger than the distance between
two filters; as a consequence ${\cal N}\gg 1$. This is the usual situation
which applies when the source parameters are not known in advance.
However, about a thousand NS's are known today -- mainly in the 
radio band -- and in looking for GW's emitted by such objects the size 
of ${\cal P}$ is drastically reduced, as some or all of the parameters 
are known a priori  with a fair degree of accuracy.

We define:
\be
\lambda_{\rm min}^j \le \lambda^j \le \lambda_{\rm max}^j
\quad\quad j = 1,..,N
\label{err}
\ee
the parameter range over which one needs to carry out a search, so that
the "error bar" is:
\be
\Delta \lambda^j  = \lambda_{\rm max}^j - \lambda_{\rm min}^j
\quad\quad j = 1,2,..,N\,.
\label{err1}
\ee
We distinguish then two cases:
\begin{itemize}
\item {\it Blind searches}: here $\lambda_{\rm max}^j \gg \lambda_{\rm min}^j$, 
and $\Delta \lambda^j \sim \lambda_{\rm max}^j $; $\lambda_{\rm max}^j$ and
$\lambda_{\rm min}^j$ are decided by the observer, based on the available 
theoretical understanding of the astrophysical scenario;
\item {\it Targeted searches}: $\lambda_{\rm max}^j \sim \lambda_{\rm min}^j$, and
$\Delta \lambda^j/ \lambda^j < 1$, or even $\ll 1$; $\lambda_{\rm max}^j$ and
$\lambda_{\rm min}^j$ are obtained from electromagnetic observations of the source.  
\end{itemize}
In targeted searches, the analysis presented in the previous section
for the estimation of the number of parameters cannot be applied
directly. It must be preceded by the determination of the actual
search parameters -- which ones and how many -- in other words
the number of dimensions of the signal manifold. In fact, some of the parameters,
might be known so accurately  -- where 
the required accuracy is determined by the loss of SNR that cannot
exceed the maximum mismatch $\mu$ -- that one does not need to
search through these parameters at all. Of particular interest
is to determine whether any known NS is a {\it one-filter target}, that 
is ${\cal N} = 1$. If this condition is satisfied, we need just 
one filter, constructed with the parameter values provided by the
available observations. 

It is easy to provide an intuitive example that shows the danger
of applying the analysis of Sec.~\ref{subs:nfilt}, without first checking
the effective number of dimensions of the signal manifold.
Assume, for sake of simplicity, that a generic GW signal $h$
depends only on two parameters, $\lambda^1$ and $\lambda^2$, 
with errors given by Eqs.~(\ref{err}) and~(\ref{err1}); 
the metric $\gamma_{jk}$, associated with this particular waveform, is 
represented by a $2\times 2$ matrix. The number of filters given by 
Eqs.~(\ref{vp}) and~(\ref{Nf}) is ${\cal N}(\Delta\lambda^1, \Delta\lambda^2$), 
where we have explicitly included the size of the parameter space in the 
argument of ${\cal N}$. It is clear that by making, say, $\Delta \lambda^2$ 
small enough, we can always obtain ${\cal N} < 1$. In other words our uncertainty
on $\lambda^1$ can be arbitrary large, but it seems that by refining our knowledge
on $\lambda^2$ we can search for that signal with only one filter.
This conclusion is clearly wrong; the correct physical interpretation 
is that, $\lambda^2$ is not a search parameter, and we can assign
its best value, provided by the available observations, to the entire bank of
filters. However, we still need to search over $\lambda^1$, which will give us 
the requisite bank of filters. The key point is therefore to decide what is the 
effective number of dimensions -- how many parameters one needs
to search for -- and then apply the analysis of the previous section to
search through the appropriate parameter space. 

Given any signal model $h(t;\l)$ and the errors in the parameters
(\ref{err}),~(\ref{err1}), we provide here a criterion to decide the effective
number of dimensions. We consider the generic N-dimensional parameter space 
described by the metric $\gamma_{jk}$, Eq.~(\ref{gamma}), 
where $j,k = 1,2,...,N$. 
The proper distance $dl$ between two filters is given by Eq.~(\ref{dl}).
The key quantity is the "{\it thickness}" $\cT^j$ of the parameter space
along any particular direction $\lambda^j$ associated with $\Delta \lambda^j$. 
From standard geometrical analysis we can derive it in a straightforward way:
\be
\cT^{j} = \frac{\Delta \lambda^j}{\sqrt{\gamma^{jj}}}\,.
\label{tau}
\ee
Here $\gamma^{jk}$ is the inverse of $\gamma_{jk}$, which is related to
$\tilde\gamma^{jk}$ by $\gamma^{jk} = \tilde\gamma^{jk}/\chi^2$;
$\sqrt{\gamma^{jj}}$ is the length of the unit vector orthogonal to the 
hypersurface $\lambda^j = $ const. The condition such that 
a particular $\lambda^j$ is not a search parameter is
\be
\max_{\gras{\lambda}\in{\cal P}}\left[\cT^{j}\right] \ll dl\,,
\label{thick}
\ee
which, via Eqs.~(\ref{dl}) and~(\ref{tau}), becomes:
\be
\Delta \lambda^j \ll 2\, \sqrt{\frac{\mu}{N}}\,
\max_{\gras{\lambda}\in{\cal P}}\left[\sqrt{\gamma^{jj}}\right]\,.
\label{thick1}
\ee
If the previous inequality yields for all the $N$ parameters,
the source needs at most a few filters. More rigorously, for a source to be exactly a 
one-filter target, the {\it diameter} of the parameter space, namely, the 
maximum distance between any pair of points in ${\cal P}$, should be less 
than $dl$. The parameter space is now a $N$-dimensional parallelepiped with 
$2^N$ vertices. This yields the following condition on the parameter errors:
\be
\max \sqrt {\gamma_{jk} \Delta \lambda^j \Delta \lambda^k } < 2 \sqrt{ \mu \over N}\,,
\label{ppd}
\ee 
where the maximum is taken over diametrically opposite $2^{N-1}$ pairs of 
vertices of the parallelepiped.  If Eq.~(\ref{ppd}) is satisfied, 
${\cal N} = 1$ and
the template is constructed by setting the parameters equal to their best fit
provided by the observations at hand. In Eq.~(\ref{ppd}) we have tacitly assumed  
that the size of the parallelepiped is smaller than the scale on which the manifold 
curves.

If $M$ $(\le N)$ parameters 
satisfy the condition~(\ref{thick}), or, equivalently, (\ref{thick1}), 
we call $\bar{\gamma}_{jk}$ the $(N-M)\times (N-M)$ metric that describes the problem,
which is constructed using the $(N-M)$ parameters that do not satisfy
Eq.~(\ref{thick1}) in the very same way described in
the previous section. The signal manifold is of lower dimension,
and the total number of filters required can be computed using 
Eqs.~(\ref{vp}) and~(\ref{Nf}) for the relevant number of dimensions $(N-M)$:
\be
{\cal N}_{N-M} \simeq \rho_{N-M} (\mu) \int_{{\cal P}_{N-M}} dV_{N-M}\,
\sqrt{\det ||\bar{\gamma}_{jk}||}\,.
\label{Nf_NM}
\ee
Here the index $(N-M)$ in the density, the number of filters,
parameter space, and volume, stresses the fact that we are considering 
a $N-M$ dimensional manifold; we also use $\simeq$ instead of "equal", because of 
possible edge effects, introduced by the dimensions which are close to, but not 
exactly, known. In the following we discuss more in detail this issue, and the 
subtleties involved in the determination of the number of parameters.

The picture becomes complicated in the case where 
the thickness of the parameter space in one or more dimensions
is smaller than, but of the same order of, the proper distance 
between two filters: $\cT^j \simlt dl$. In this case it is useful
to introduce the following notation:
\be
\cT^j = \delta\,dl\quad\quad 0 < \delta \le 1\,.
\label{thick2}
\ee
When $\delta \sim 1$, we must be able to reach the signal on the boundary
of the parameter space,
which means that the filters should be placed closer than in the case 
of $\delta \ll 1$. In the general case, the correct determination of the
number of filters can be very complicated, due to the edge effects. For
sake of simplicity, let us consider, first, the case where only one
parameter, say $j = 1$, satisfies Eq.~(\ref{thick2}). In this situation, we have,
\be
(N - 1) \left(\frac{dl}{2}\right )^2 + \left (\frac{\cT^1}{2}\right)^2 = \mu\,.
\label{thick2a}
\ee
Solving the above equation we have,
\be
dl = 2\, \sqrt{\frac{\mu}{N - 1}\,\left(1 - \frac{\delta^2}{N}\right)},
\label{dl1}
\ee
and since the unit of volume in parameter space is $dV_{\rm patch} = dl^{N-1}$, 
the total number of filters is given by
\be
{\cal N} = \rho_{N-1} (\mu) \int_{{\cal P}_{N-1}} dV_{N-1}\, 
\sqrt{\left(1 - \frac{\delta^2}{N}\right)^{(1 - N)}\,
\det ||\bar{\gamma}_{jk}||}\,,
\label{Nf1}
\ee
where $j,k = 2,..,N$. It is easy to check that we recover the expected 
formulae for the parameter space volume from Eq.~(\ref{Nf1}) in the limiting 
cases when $\delta = 0, 1$.  

In more general cases, when two or more parameters are characterized by
$\delta \simlt 1$, the situation becomes very complex, and an "exact"
determination of the number of filters particularly hard. In the
case when the signal manifold is 3-dimensional and two parameters, say 
$\lambda_1$ and $\lambda_2$,
are known with $\delta \simlt 1$ it is still possible to obtain a
rather simple expression for the thickness. In this case:
\be
\cT = \left\{\sum_{j,k=1}^2 
\left[\left(\gamma_{jk} - \frac{\gamma_{j3}\gamma_{k3}}{\gamma_{33}}\right)
\Delta\lambda^j\Delta\lambda^k\right]\right\}^{1/2}
\label{thick12}
\ee
A comparison analogous to the previous one then applies.

It is however possible to obtain an estimate of the total number of
filters which is correct within a factor of a few in the general case
(we will use this criterion in Sec.~\ref{sec:obs}, when we estimate
the computational costs for known radio pulsars).
We introduce the following notation: $M$ $(\le N)$ is the
number of parameters that satisfy Eq.~(\ref{thick2}), and $M' (\le M)$ 
those for which $\delta \ll 1$; the $M'$ parameters satisfy
Eq.~(\ref{thick}) or~(\ref{thick1}), and in practice we can
consider them as "exactly" known; the value of $\delta$ that we
adopt to distinguish the $M'$ parameters from the other $M-M'$ ones
is $\delta = 0.1$. Indeed, $M-M'$ is the number of parameters for 
$0.1 < \delta \le 1$, those for which edge effects may become important. 
The reason of the choice $\delta = 0.1$ is the following: in the
case $M' = M = 1$, it is easy to verify, through Eq.~(\ref{Nf1}), that 
the estimate of the total number of filters is very accurate. We assume
that similar accuracy will hold also for few  number of dimensions.
The values $N$, $M$ and $M'$ satisfy therefore the inequality
$M'\le M \le N$. To summarize, we consider the $M'$ parameters to be
exactly known, $M-M'$ "questionable" (border effects need to be 
taken into account) and $N-M$ to search for. One can imagine the questionable 
parameters forming a $M-M'$ dimensional parallelepiped at each point of the 
$N-M$ dimensional sub-manifold. The projection of this parallelepiped `orthogonal' 
to the $N-M$ dimensional sub-manifold of search parameters decides the number 
of filters needed for the questionable parameters. The total number of 
filters is therefore:
\be
{\cal N} \simlt 2^{(M-M')}\, {\cal N}_{(N-M)} ,
\label{Nf2}
\ee
where ${\cal N}_{(N-M)}$ is given by Eq.~(\ref{Nf_NM}). The
multiplicative factor $2^{(M-M')}$ accounts for the number of filters (which we
have overestimated) to take care of each of the questionable parameters. Here, we 
have been cautious to assume  two filters per parameter,
say one corresponding to $\lambda_{\rm min}^j$ and one to
$\lambda_{\rm max}^j$. It is simple to check
that we overestimate the number of filters by applying Eq.~(\ref{Nf1}),
which is valid if $M' = 0$ and $M = 1$. In the case $M - M'\gg 1$ Eq.~(\ref{Nf2})
might greatly overestimate the actual number of filters. In practical
cases $M-M'$ is expected to be at most a few, so that the expression~(\ref{Nf2})
should provide a meaningful upper-limit (clearly ${\cal N}_{N-M} < {\cal N}
< 2^{(M-M')} {\cal N}_{(N-M)}$); however, one can easily resolve this issue
by setting up a Monte-Carlo experiment to determine the additional number of 
filters and their location in ${\cal P}$ for the questionable dimensions.

%
%

\section{The computational costs}
\label{sec:cost}

\subsection{The signal phase and dimensionless parameters}
\label{subs:phase}

It is convenient for the purposes of computation to express the phase 
$\Phi(t)$, Eq.~(\ref{phiT}), in terms of dimensionless parameters 
and use the above mentioned scaling properties, 
Eqs.~(\ref{sc_phi}) to (\ref{sc_v}), 
to obtain the actual proper volume $\Vp$, Eq.~(\ref{vp}). We write
\be
\Phi = (2 \pi f_0 T)\, \tPhi,
\label{phase1}
\ee
where
\be
\tPhi = \kappa u + X \cos E  + Y \sqrt{1 - e^2} \sin E\, ,
\label{tph}
\ee
and the dimensionless parameters are given by:
\bea
u & \equiv & \frac{t}{T}\,, \\
\kappa & \equiv & \frac{f - f_0}{f_0}\,, \\
X & \equiv & - a \sin \epsilon \cos \psi / c T \,,\\ 
Y & \equiv & - a \sin \epsilon \sin \psi / c T\,, \\
\Omega & \equiv & \omega T\,. 
\label{pad}
\eea
Here $u$ is a dimensionless time satisfying $0 \le u \le 1$.
So now the new set of parameters is ${\bf \Lambda} = (\kappa\,, X\,, Y\,, e\,, \Omega\,, \a)$
and $\gras{\lambda} = (X\,, Y\,, e\,, \Omega\,, \a)$. The scaling factor $\chi$ in 
Eqs.~(\ref{sc_g}),~(\ref{sc_d}) and~(\ref{sc_v}) is clearly:
\be
\chi = 2\pi\fM T\,,
\label{chi}
\ee
where we have set $f_0 = \fM$, the maximum source frequency that one searches for.
Using an equation 
analogous to Eq.~(\ref{metric}) we obtain the scaled metric $\tilde g_{\a \b}$.

The exact expression of the determinant of the metric from which we compute the 
volume is quite complicated. However, in this section our goal is to obtain 
approximate analytical expressions -- which turn out to be very 
accurate over most of the relevant parameter range -- which give us the 
important information about how the number of filters scales as a function of 
the key observables. For this purpose it is necessary to 
make some assumptions about the ranges of these parameters from 
astrophysical scenarios. The semi-major axis $a_{\rm NS}$ of a NS orbiting a companion
of mass $m_2$ -- with mass ratio $q \equiv m_2/m_{\rm NS}$ -- is:
\be
a_{\rm NS} \simeq 2.65\times 10^{11}\,\left(\frac{m_{\rm NS}}{1.4\,\Ms}\right)^{1/3}\,
\left(\frac{\omega}{10^{-4}\,{\rm rad}/{\rm sec}}\right)^{-2/3}\,
\left[\frac{q}{(1 + q)^{2/3}}\right]\,{\rm cm}
\label{aNS}
\ee
where the choice of $\omega$ corresponds to  $\simeq 17$ hours orbital 
period (notice that
for a NS in a 2 hours orbit, the semi-major axis would be a factor 
$\simeq 4.2$ smaller). We take
the time of observation to be typically 4 months long, so that $T = 10^7$ sec and
we set, as reference, $\fM = 1$ kHz. Thus the factor $\fM T \sim 10^{10}$. The 
dimensionless parameters $X$ and $Y$ are $X\sim Y\sim a_{\rm NS}/cT$, and
give essentially the size of the orbit in units of the distance light (or actually GW) 
travels during the observational time $T$. They are related to $\Omega$ by:
\be
\frac{a_{\rm NS}}{c\,T} \simeq 8.8\times 10^{-7}\,
\left(\frac{m_{\rm NS}}{1.4\,\Ms}\right)^{1/3}\,
\left(\frac{\Omega}{10^3}\right)^{-2/3}\,\left(\frac{T}{10^7\,{\rm sec}}\right)^{-1/3}\,
\left[\frac{q}{(1 + q)^{2/3}}\right]\,,
\ee
\be
\Omega = 10^3 \left(\frac{\omega}{10^{-4}\,{\rm rad}/{\rm sec}}\right)\,
\left(\frac{T}{10^7\,{\rm sec}}\right)\,.
\label{Om}
\ee
Notice that the value of $\Omega$ corresponds to  
an observation that extends over several orbits of the binary system. In fact
$\Omega$ is the number of radians, or $\Omega /2 \pi$ is the number of orbits,
that the star completes in the time $T$. In this case, one can Taylor expand
the matrix elements, and therefore the determinant, as a function of 
$\Omega^{-1} \ll 1$. Note, moreover, that in this case $X \sim Y \ll \Omega^{-1} \ll 1$, that is $a$ and $\omega$ become of the same order -- in geometrical units,
$c=G=1$ -- only for a NS orbiting a companion object at a distance of the 
order of the gravitational radius. We will use these results in the approximation
scheme, retaining only the leading order terms in $X$ and $Y$.

The other physically relevant case is in the opposite limit $\Omega \ll 1$. 
We therefore Taylor expand the matrix determinant as a function of 
$\Omega$. This case applies to observations that cover only a fraction of one 
orbital period, and the closed form expression that we obtain turns out to 
provide a rather good approximation of the full expression upto 
$\Omega \simeq 1$.    

In the next subsection we examine the case when the orbit is 
circular, that is, $e = 0$. We treat this simpler case first in order to obtain 
useful insights, before we address the general case of the elliptical orbit.

\subsection{Circular orbits}
\label{subs:circ}

The circular case is important both from the pedagogical and physical point of
view: (a) it provides us several insights into the problem
via a comparatively easier computation; (b) in targeted searches, several
known NS's in binary systems, including Sco X-1, are 
essentially in a circular orbit; (c) for blind searches, present and near future
processing power is likely to allow us to search over a reasonable parameter space
mainly for emitters orbiting a companion with $e = 0$.

For circular orbits, the expression of the phase~(\ref{tph}) simplifies considerably:
here $e = 0$, and $\psi$ and $\a$ combine additively into a single parameter which 
we redefine again as $\a$, for sake of simplicity; in effect we put $\psi = 0$. Then 
$X$ is just the projected radius of the orbit which we denote by $A$. We therefore 
have just 3 search parameters for which a discrete mesh of filters is required: 
${\bf \Lambda} = (\kappa\,,A\,,\Omega\,, \a)$ and ${\gras \lambda} = (A\,,\Omega\,, \a)$.
The phase~(\ref{tph}) becomes therefore:
\be
\tPhi = \kappa u + A \cos (\Omega u + \a)\,.
\label{circphs}
\ee
The metric $\tg_{\a \b}$, see Eq.~(\ref{metric}) is now a $4 \times 4$ matrix,
and we compute it from Eq.~(\ref{metric}) 
-- with quantities scaled by the factor $2\,\pi\,f_0\,T$, and therefore
replaced with `tilde' -- and Eq.~(\ref{circphs}). 
The first derivatives are simply:
\bea
\tPhi_{\kappa} & = & u \,,\nonumber\\
\tPhi_{A} & = & \cos (\Omega u + \a)\,,\nonumber\\
\tPhi_{\Omega} & = & - A u \sin (\Omega u + \a)\,,\nonumber\\
\tPhi_{\a} & = & - A \sin (\Omega u + \a)\,.
\label{circder}
\eea
We need to compute the 10 time averages involving the above $\tPhi_\beta$ 
($\beta = 0,1,2,3$)
and then compute the metric $\tgamma_{jk}$ through the projection of $\tg_{\a \b}$
orthogonal to $\kappa$, as discussed in Sec.~\ref{sec:datan}. The exact analytical 
expression
of $\tVp = \sqrt{\det{||\tgamma_{jk}||}}$ is very complex and not very illuminating. 
However, as discussed in the previous section, it is possible to compute 
$\det{||\tgamma_{jk}||}$ in a closed form in the two relevant regimes:
(i) $\Omega \gg 1$, the limit of
several orbits during the observation time $T$, and (ii) 
$\Omega \ll 1$, the limit of monitoring a fraction of an orbit. 

\subsubsection{The limit of large number of orbits}

\begin{figure}
\begin{center}
\epsfig{file=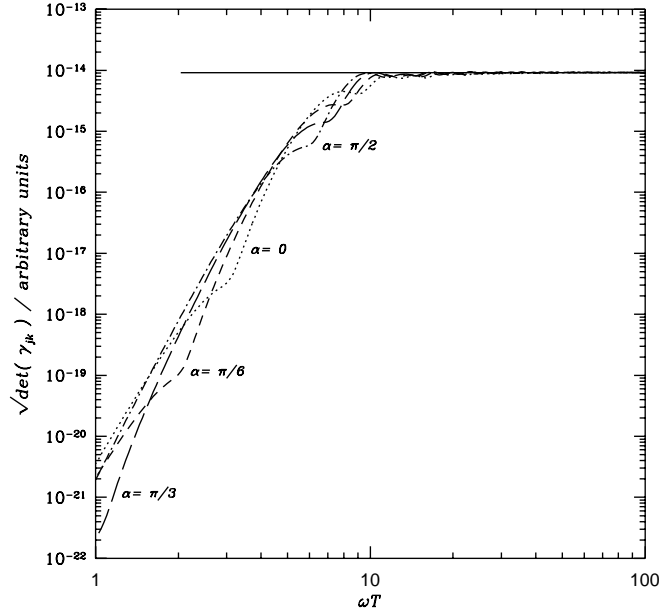,width=4.1in}
\caption{\label{fig:circgg}
Comparison of the asymptotic analytical expression and the numerical evaluation
of the full determinant in the large number of orbits approximation $(\Omega \gg 1)$,
for sources in circular orbits. 
The plot shows the proper volume element $\sqrt{\det{||\gamma_{jk}}||}$,
in arbitrary units, as a function of the dimensionless orbital frequency parameter
$\Omega \equiv \omega T$. The solid line corresponds to the asymptotic expression
in the regime $\Omega \gg 1$, Eq.~(\protect{\ref{detc1}}),
whereas the other curves correspond to the numerical
evaluation of the full expression of $\sqrt{\det{||\gamma_{jk}}||}$, for several
values of the initial phase of the orbit of the source: $\alpha = 0$ (dotted line),
$\alpha = \pi/6$ (dashed line), $\alpha = \pi/3$ (long-dashed line), and
$\alpha = \pi/2$ (dotted-dashed line). The size of the projected semi-major axis
in dimensionless units is $A = 10^{-7}$. Notice that for $\Omega \ge 20$ the
asymptotic expansion is indistinguishable from the full expression.
}
\end{center}
\end{figure}

We derive the expression of the determinant of $\tgamma_{ij}$ in the limit
$\Omega \gg 1$. The elements of $\tg_{\a \b}$
and $\tgamma_{jk}$ are explicitly given in Appendix~\ref{app:e0gg1}. 
The proper volume element reads:
\be
\sqrt{\det{||\tgamma_{jk}}||} = {A^2 \over \sqrt{96}} + o(\Omega^{-1}, A^3)\,.
\label{detc1}
\ee
Fig.~\ref{fig:circgg} shows the comparison of the analytical asymptotic 
expression~(\ref{detc1}) with the numerical evaluation of the full
determinant. It is remarkable that for $\Omega \simgt 10$ -- which 
corresponds to about 2 orbits completed during $T$ -- the two
results are essentially identical.
 
The proper volume $\tVp$ is given by integrating the volume element~(\ref{detc1}) 
within  the appropriate limits. Here we need to integrate over 
$A, \Omega$ and $\a$, within the ranges:
\bea 
A_{\rm min} & \le & A \le A_{\rm max}\,, \nonumber\\
\Omega_{\rm min} & \le & \Omega \le \Omega_{\rm max}\,, \nonumber\\
0 & \le & \a \le 2 \pi\,.
\label{limc}
\eea 
Inserting Eq.~(\ref{detc1}) in Eq.~(\ref{vp}), and taking into
account the limits~(\ref{limc}), we get the following proper scaled volume:
\be
\tVp = {\pi \over 6 \sqrt{6}}\,(A_{\rm max}^3 - A_{\rm min}^3)
\,(\Omega_{\rm max} - \Omega_{\rm min})\,.
\label{vpsc1}
\ee
Since the parameter space is 3-dimensional, $N = 3$, 
to obtain the actual volume we need to multiply 
$\tVp$ by the factor $\chi^3 = (2 \pi f_{\rm max} T)^3$:
\be
\Vp = {\pi \over 6 \sqrt{6}}\,\left(\frac{2\pi f_{\rm max}}{c}\right)^3\, [\apM^3 
- \apm^3 ] (\oM - \om ) T\, .
\label{vpc1}
\ee
Notice that the factor $2\pi f_{\rm max}/c$ is the maximum wavenumber 
of the gravitational wave that we want to detect. The main point in Eq.~(\ref{vpc1})
is to observe how the volume scales. The number of 
filters increases linearly in the observation time $T$, so hierarchical searches  
based on $T$ will not work effectively (compare this to the case of the all sky all frequency 
searches for isolated pulsars where the patches scale as $T^5$). The volume is 
proportional to the cube of the size of the projected orbit along the line of 
sight $a_{\rm p}$. So knowing, say, the inclination angle fairly well and 
the radius of the orbit will greatly reduce the computational load. Similar
considerations apply to the frequency.

In the case of blind searches, $\Delta \lambda^j \sim \lambda^j_{\rm max}$, 
for all the parameters. Then the parameter volume is:
\be
\Vp \simeq 1.96 \times 10^{15}\,\left(\frac{f_{\rm max}}{1\,{\rm kHz}}\right)^3
\,\left(\frac{\apM}{10^{11}\,{\rm cm}}\right)^3\,
\left(\frac{\oM}{10^{-4}\,{\rm rad/sec}}\right)\,
\left(\frac{\alM}{2\pi}\right)
\left(\frac{T}{10^{7}\,{\rm sec}}\right);
\label{Vbc1}
\ee
and the total number of filters is given by:
\be
{\cal N}(\mu = 3\%) \simeq 1.25\times 10^{17}\,\left(\frac{\Vp}{10^{15}}\right)\,.
\label{Nfc_gg}
\ee
This result indicates that the processing power to carry out a coherent
blind search (even in the case of circular orbits) is outrageous; in fact,
assume that the the costs are entirely dominated by the computation
of the FFT and the power spectrum; then, each filter, over the entire search
bandwidth, requires the following number of floating point operations:
\bea
n_{\rm op} & = & 6 \fM T \left[\log_2(2 \fM T) + \frac{1}{2}\right] \nonumber\\
& \simeq & 2.1 \times 10^{12} \left(\frac{\fM}{1\,{\rm kHz}}\right)\,
\left(\frac{T}{10^7\,{\rm sec}}\right)\,.
\label{nop}
\eea
As a consequence the total processing power or computational speed $\S$ to 
keep up with the data is:
\bea
\S & = & \frac{n_{\rm op} {\cal N}}{T} \nonumber\\
& \simeq & 2.1 \times 10^{16}\,
\left(\frac{\fM}{1\,{\rm kHz}}\right)\,
\left(\frac{{\cal N}}{10^{17}}\right)\,{\rm MFlops}\,.
\label{power}
\eea
We would like to stress that this estimation applies only to the orbital
parameters and does not take into account the costs involved in searching for the 
source location in the sky and the spin down parameters. We would also like
to point out the very steep dependence of the processing power, cfr. Eq.~(\ref{Vbc1})
and~(\ref{power}), on the maximum frequency $\fM$ upto which the search is 
conducted: $\S \propto \fM^4$. This indicates that restricting the
frequency band translates into a major saving in computational power.
As one could have imagined, from previous results pertaining to isolated sources,
the additional costs to search for the orbital parameters are absolutely
prohibitive. This is also due to the fact, that the acceleration of a source
with an orbital companion may be much higher than detector acceleration 
due to Earth's rotation and orbital motion, producing a much larger frequency 
drift which needs to be corrected for. Clearly, only prior information
on the source parameters can make such a search feasible. In fact, when some
parameters are known a priori, the number of templates
to process the data reduces drastically. 

\begin{figure}
\begin{center}
\epsfig{file=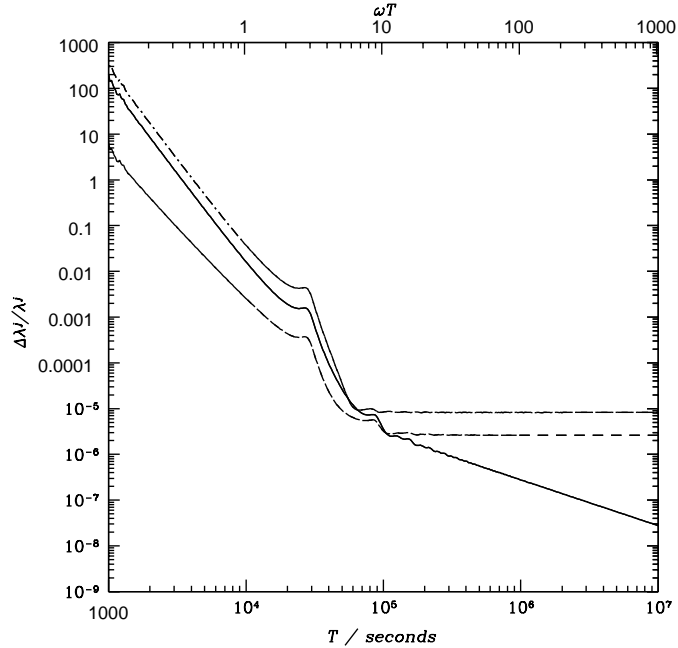,width=4.1in}
\caption{\label{fig:err0}
The errors on the orbital parameters required to carry out
a search using only a few matched filters. The
relative errors in the parameters $\Delta \omega/\omega$ (solid line), 
$\Delta a_{\rm p}/a_{\rm p}$ (dotted-dashed line ) and 
$\Delta\alpha/\alpha$ (dashed line) are plotted for a constant thickness 
of $\cT^j = dl = 0.2$ ($N = 3$ and $\mu = 0.03$); the source corresponds to
a NS/NS binary system in circular orbit; the parameters have been chosen
to have the following values:
$\omega = 1.027\times 10^{-4}$ rad$/$sec, $a_{\rm p} = 1.64\times 10^{11}$ cm,
$\alpha = 0.1$ rad, and $\fM = 1$ kHz.
}
\end{center}
\end{figure}

It is interesting to estimate the thickness of the parameter
space along the different directions as a function of the 
parameter values and errors, in order to gain insight into the
accuracy that is required to carry out a search only with a few
templates. For a 3 parameter search, and a mismatch $\mu = 3\%$
between the signal and the template, the filter separation, see
Eq.~(\ref{dl}), is $dl = 0.2$; it  has to be compared with the thickness $\cT^j$,
Eq.~(\ref{thick}), for $j = 1,2,3$. Using our approximate
expression of the metric, cfr. Appendix~\ref{app:e0gg1}, we obtain:
\bea
\cT^{a_{\rm p}} & \simeq & 1.49\times 10^{-2}\,
\left(\frac{\Delta a_{\rm p}}{1\,{\rm km}}\right)\,
\left(\frac{f_{\rm max}}{1\,{\rm kHz}}\right)\,,
\nonumber\\
\cT^{\omega} & \simeq & 4.29\times 10^{-2}\,
\left(\frac{\Delta\omega}{10^{-12}\,{\rm sec}^{-1}}\right)\,
\left(\frac{a_{\rm p}}{10^{11}\,{\rm cm}}\right)\,
\left(\frac{T}{10^7\,{\rm sec}}\right)\,
\left(\frac{f_{\rm max}}{1\,{\rm kHz}}\right)\,,
\nonumber\\
\cT^{\alpha} & \simeq & 7.43\times 10^{-2}\,
\left(\frac{\Delta\alpha}{10^{-5}\,{\rm rad}}\right)\,
\left(\frac{a_{\rm p}}{10^{11}\,{\rm cm}}\right)\,
\left(\frac{f_{\rm max}}{1\,{\rm kHz}}\right)\,.
\label{thick_cgg1}
\eea
For typical binary neutron stars parameters, we also show in Fig.~\ref{fig:err0} 
the relative errors $\Delta\lambda^j/\lambda^j$ that are required so
that $\cT^j < dl$. Our prior knowledge on the parameters must be very accurate
in order to reduce the number of filters to only a few; typically an
error $\Delta\lambda^j/\lambda^j < 10^{-5}$ (or smaller) is required. 
However, this is not at all uncommon for radio observations and this aspect   
will be discussed in Section~\ref{sec:obs}.

\subsubsection{The fraction of one orbit case}

We consider now the opposite limit: $\Omega \ll 1$. This corresponds to the
physical situation where pulsar radio-astronomers usually apply the
so-called "accelerated search"\cite{AGKPW90}, as one monitors only a small 
fraction of the source orbital period, say less than one radian. The presence of a 
companion object for such a short time introduces an acceleration 
on the source motion that could be, as first approximation, treated as constant.

For GW observations, the case $\Omega \ll 1$ is relevant for  
binary systems with very long orbital periods (of the order of one year or more)
and $T \sim 10^7$ sec, or when coherent integrations are performed
on short data segments (say of the order of one hour). The latter situation
is often encountered in hierarchical data analysis schemes,
where one first uses matched filters on short time base-lines, and
then concatenates together in a incoherent fashion the corrected data chunks.

We derive an approximate expression for the determinant of the metric 
$\tgamma_{jk}$ by Taylor expanding it as a function of $\Omega \ll 1$. 
The computation is trivial but cumbersome, and we
present the explicit expressions of the elements of the reduced metric, 
Eq.~(\ref{sc_v}), with phase model~(\ref{phase1}) in the Appendix~\ref{app:e0ll1}. 
We obtain the following volume element:
\be
\sqrt{\det ||\gamma_{jk} ||} =
\frac{\left(1 + \cos 2\alpha\right)^{1/2}}{3628800\sqrt{70}}\,
A^2\,\Omega^8 + o(\Omega^9, A^3)\,.
\label{detc2}
\ee
It is interesting to note that for $\alpha = \pi/2$ and $3\pi/2$ -- the NS is 
receding from, or approaching toward the detector essentially along the  
line of sight -- the determinant tends to zero (at this order, but 
there are higher order corrections in $\Omega$ that depend on $\alpha$,
and do not vanish): in fact the signal appears Doppler-shifted by a constant 
indistinguishable off-set, depending on the velocity of the  source, that one
does not need to correct for. 

\begin{figure}
\begin{center}
\epsfig{file=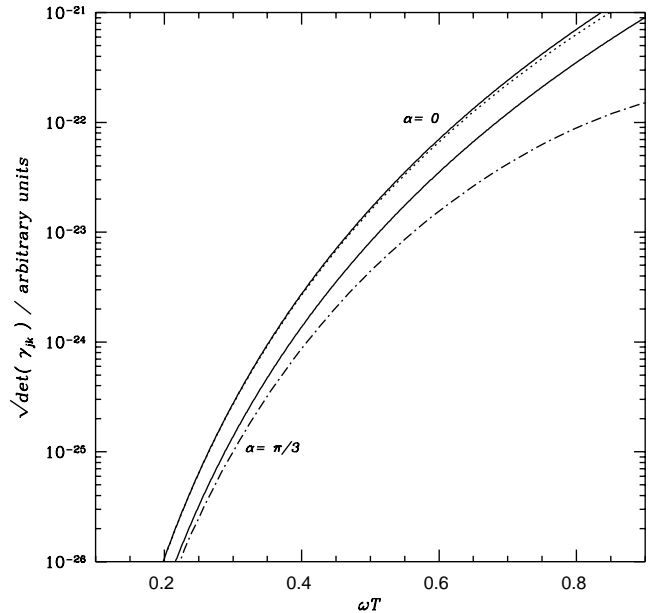,width=4.1in}
\caption{\label{fig:circll1}
Comparison of the asymptotic analytical expression and the numerical evaluation
of the full determinant for the case of observations lasting a fraction of 
the neutron-star orbit ($\Omega \ll 1$). The plot shows the proper volume element 
$\sqrt{\det{||\gamma_{jk}}||}$, in arbitrary units, as a function of the 
angular orbital frequency, in units of the observation time, $\Omega \equiv \omega T$. 
The solid lines correspond to the asymptotic expressions
in the regime $\Omega \simlt 1$, see Eq.~(\protect{\ref{detc2}}),
for selected values of the initial orbital phase
of the source: $\alpha = 0$ and $\pi/3$ (see labels). For the same values of
$\alpha$ we plot also $\sqrt{\det{||\gamma_{jk}}||}$ obtained by numerically
evaluating the full expression (dotted line: $\alpha = 0$; dashed-dotted line:
$\alpha = \pi/3$). 
}
\end{center}
\end{figure}

A comparison of the asymptotic limit~(\ref{detc2}) with the full expression of 
the determinant is shown in Fig.~\ref{fig:circll1}, where we present the
results for two values of $\alpha = 0\,,\pi/3$. Notice that Eq.~(\ref{detc2})
is very accurate upto $\Omega \simeq 1$ for $\alpha = 0$, whereas it approximates 
correctly the full expression only for $\Omega \simlt 0.1$ in the case $\alpha = \pi/3$. 
This is due to the
fact that when $\alpha \ne 0$, the determinant contains higher order correction
terms in $\Omega$, that depend on $\alpha$ and vanish for $\alpha = 0$. In the
range $0.1 \simlt \Omega \simlt 1$ they produce a sizable effect that one cannot ignore.

By integrating Eq.~(\ref{detc2}) on the parameter range~(\ref{limc}), 
and following steps similar to the previous case, we obtain the volume:
\be
\Vp = \frac{1}{24494400\sqrt{35}}\,
\,\left(\frac{2 \pi \fM}{c}\right)^3\, [\apM^3 
- \apm^3 ]\,\left[(\oM T)^9 - (\om T)^9\right]\,.
\ee
Following the same scheme as in the previous section, we can obtain
the total volume of the parameter space, when no prior
information about the source parameters is available:
\bea
\Vp & \simeq & 6.3 \,\left(\frac{f_{\rm max}}{1\,{\rm kHz}}\right)^3
\,\left(\frac{\apM}{10^{11}\,{\rm cm}}\right)^3\,
\left(\frac{\oM}{10^{-4}\,{\rm rad/sec}}\right)^9\,
\left(\frac{T}{1\,{\rm hr}}\right)^9\nonumber\\
& \simeq & 3.3\times 10^5\,\left(\frac{f_{\rm max}}{1\,{\rm kHz}}\right)^3
\,\left(\frac{\apM}{10^{13}\,{\rm cm}}\right)^3\,
\left(\frac{\oM}{10^{-7}\,{\rm rad/sec}}\right)^9\,
\left(\frac{T}{1\,{\rm month}}\right)^9\,.
\eea
The total number of filters, for a mismatch of $3\%$ and the parameters 
values quoted in the previous expression is therefore ${\cal N} \simeq 793$
and $4.2\times 10^7$ respectively. The extra-computational
burden to carry out searches of NS in binaries with long orbital periods 
as compared with the observation time, might be therefore affordable or
even negligible in some cases. In fact, the processing power required
to keep up with the data flow is respectively:
\bea
\S & \simeq & 1.4\times 10^2 
\left(\frac{\fM}{1\,{\rm kHz}}\right)\,
\left(\frac{{\cal N}}{10^{3}}\right)\,
{\rm MFlops}\,,\nonumber\\
& \simeq & 1.9\times 10^6
\left(\frac{\fM}{1\,{\rm kHz}}\right)\,
\left(\frac{{\cal N}}{10^{7}}\right)\,
{\rm MFlops}\,.
\label{power1}
\eea

The costs are of course reduced if some orbital elements are known in advance, and
the parameters need to be known only within a factor $\approx 2$ to allow
a few templates search. In fact, the thickness of the parameter space in 
the three directions becomes now:
\bea
\label{thickc_ll1}
%
%
\cT^{a_{\rm p}} & \simeq & 
\left\{ \begin{array}{ll}
1.96\times 10^{-2}\,
\left(\frac{\sqrt{1 + \cos(2\alpha)}}{5 - \cos(2\alpha)}\right)\,
\left(\frac{\Delta a_{\rm p}}{10^{10}\,{\rm cm}}\right)\,
\left[\left(\frac{\omega}{10^{-4}\,{\rm sec}^{-1}}\right)\,
\left(\frac{T}{1\,{\rm hr}}\right)\right]^4\,
\left(\frac{f_{\rm max}}{1\,{\rm kHz}}\right)
& \quad\quad \left[\alpha \ne \left(k + \frac{1}{2}\right) \pi\right] \nonumber\\
%
%
7.6\times 10^{-3}\,
\left(\frac{\Delta a_{\rm p}}{10^{10}\,{\rm cm}}\right)\,
\left[\left(\frac{\omega}{10^{-4}\,{\rm sec}^{-1}}\right)\,
\left(\frac{T}{1\,{\rm hr}}\right)\right]^5\,
\left(\frac{f_{\rm max}}{1\,{\rm kHz}}\right)
& \quad\quad \left[\alpha = \left(k + \frac{1}{2}\right) \pi\right]\nonumber \\
\end{array}
\right.
\nonumber\\
%
%
\cT^{\omega} & \simeq & 
\left\{ \begin{array}{ll}
9.9\times 10^{-2}\,
\left[\sqrt{1 + \cos(2\alpha)}\right]\,
\left(\frac{\Delta\omega}{10^{-4}\,{\rm sec}^{-1}}\right)\,
\left(\frac{a_{\rm p}}{10^{11}\,{\rm cm}}\right)\,
\left(\frac{\omega}{10^{-4}\,{\rm sec}^{-1}}\right)^3\,
\left(\frac{T}{1\,{\rm hr}}\right)^4\,
\left(\frac{f_{\rm max}}{1\,{\rm kHz}}\right)
& \quad\quad \left[\alpha \ne \left(k + \frac{1}{2}\right) \pi\right] \nonumber\\
%
%
2.5\times 10^{-2}\,
\left(\frac{\Delta\omega}{10^{-4}\,{\rm sec}^{-1}}\right)\,
\left(\frac{a_{\rm p}}{10^{11}\,{\rm cm}}\right)\,
\left(\frac{\omega}{10^{-4}\,{\rm sec}^{-1}}\right)^4\,
\left(\frac{T}{1\,{\rm hr}}\right)^5\,
\left(\frac{f_{\rm max}}{1\,{\rm kHz}}\right)
& \quad\quad \left[\alpha = \left(k + \frac{1}{2}\right) \pi\right] \nonumber \\
\end{array}
\right.
\nonumber\\
%
%
\nonumber\\
\cT^{\alpha} & \simeq & 
\left\{ \begin{array}{ll}
2.8\times 10^{-2}\,
\sqrt{\frac{1 + \cos(2\alpha)}{1 - \cos(4\alpha)}}
\left(\frac{\Delta\alpha}{10^{-1}\,{\rm rad}}\right)\,
\left(\frac{a_{\rm p}}{10^{11}\,{\rm cm}}\right)\,
\left[\left(\frac{\omega}{10^{-4}\,{\rm sec}^{-1}}\right)\,
\left(\frac{T}{1\,{\rm hr}}\right)\right]^4\,
\left(\frac{f_{\rm max}}{1\,{\rm kHz}}\right)
& \quad\quad \left[\alpha \ne k\,\frac{\pi}{2} \right] \\
%
%
3.8\times 10^{-2}\,
\left(\frac{\Delta\alpha}{10^{-1}\,{\rm rad}}\right)\,
\left(\frac{a_{\rm p}}{10^{11}\,{\rm cm}}\right)\,
\left[\left(\frac{\omega}{10^{-4}\,{\rm sec}^{-1}}\right)\,
\left(\frac{T}{1\,{\rm hr}}\right)\right]^3\,
\left(\frac{f_{\rm max}}{1\,{\rm kHz}}\right)
& \quad\quad \left[\alpha = k\,\pi \right] \\
%
%
6.3\times 10^{-1}\,
\left(\frac{\Delta\alpha}{10^{-1}\,{\rm rad}}\right)\,
\left(\frac{a_{\rm p}}{10^{11}\,{\rm cm}}\right)\,
\left[\left(\frac{\omega}{10^{-4}\,{\rm sec}^{-1}}\right)\,
\left(\frac{T}{1\,{\rm hr}}\right)\right]^2\,
\left(\frac{f_{\rm max}}{1\,{\rm kHz}}\right)
& \quad\quad \left[\alpha = \left(k + \frac{1}{2}\right) \pi\right]
\end{array}
\right.
\eea
where $k$ is an integer.
The difference with respect to the limit $\Omega \gg 1$ is striking; 
Fig.~\ref{fig:err0} clearly shows the steep dependence 
of the thickness of ${\cal P}$ on $\Omega$, or, equivalently, the time of 
observation,
where the behaviour of $\Delta\lambda^j/\lambda^j$ changes abruptly when
$T \sim P$. Notice also the change of scaling of the $\cT$'s depending on 
the location of the source on the orbit, {\it i.e.} the value of the 
parameter $\alpha$.

\subsection{Elliptical orbits}
\label{subs:ell}

\begin{figure}
\begin{center}
\epsfig{file=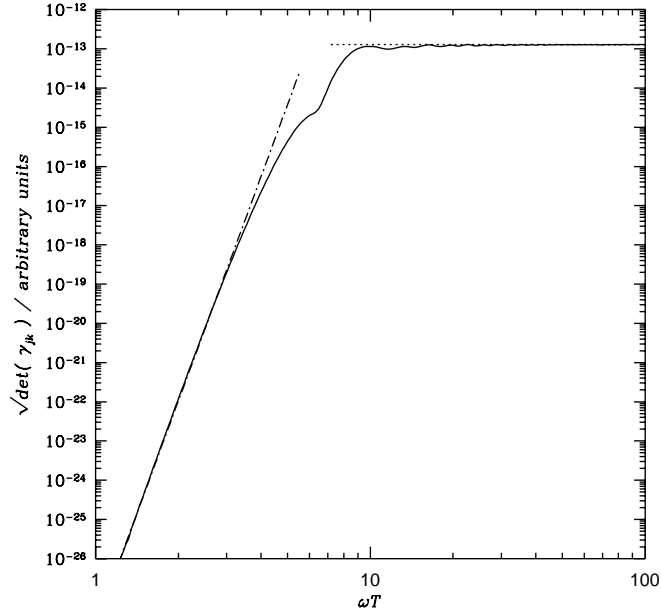,width=4.1in}
\caption{\label{fig:ell}
Comparison of the asymptotic analytical expression and the numerical evaluation
of the proper volume element of the signal manifold for neutron stars in elliptical orbits. 
The plot shows $\sqrt{\det{||\gamma_{jk}}||}$,
in arbitrary units, as a function of the dimensionless orbital frequency parameter
$\Omega \equiv \omega T$. The solid line corresponds to the exact expression of the
determinant, computed numerically. The dotted line refers to the asymptotic
expression of the determinant, cfr. Eq.~(\protect{\ref{dete}}), in the regime
$\Omega \gg 1$; the dotted-dashed line is the asymptotic expression of the
determinant for $\Omega \ll 1$, in which case 
$\sqrt{\det{||\tgamma_{jk}||}} \propto \Omega^{19}$. 
The size of the projected semi-major axis
in dimensionless units is $A = 10^{-7}$; the other orbital elements
are chosen as $e = 0.5$, and $\alpha = \psi = 0$.
}
\end{center}
\end{figure}

We have already described the orbit in Sections~\ref{subs:orbpar} and~\ref{subs:phase}. 
As phase model we use Eq.~(\ref{tph}) and take the expansion of the eccentric
anomaly $E$ as a function of the mean anomaly $M$ and the eccentricity $e$ and 
truncate at the 7th order in $e$ (see Eqs.~(\ref{orbpl}) and Appendix~\ref{app:CS}). 
This means we consider 7 harmonics in $\omega$. The volume that we obtain is 
also correct upto this order in $e$. 
The derivatives of the phase~(\ref{tph}) with respect to the six parameters
$\kappa, X, Y, e, \omega, \a$, or, equivalently, 
$\kappa, A, e, \omega, \a, \psi$ are easily computed. We then proceed as 
in the circular orbit case, except now $\tg_{\a \b}$ is a $6 \times 6$ matrix and the 
$\tgamma_{jk}$ is a $5 \times 5$ matrix. The problem is therefore much more complex 
than in the circular case, and it is impossible to obtain a closed form expression
of the determinant. However, we can still Taylor-expand the relevant expressions
as done before, and keep the leading order terms. We present here a closed  
form expression of the volume element, and therefore the number of filters, 
in the asymptotic limit $\Omega \gg 1$. The limit $\Omega \ll 1$ can in principle
be obtained by applying the same scheme adopted in the previous subsection. However
the number of terms that we need to retain is so large that the computation 
becomes very cumbersome. We present only the numerical results in Fig.~\ref{fig:ell},
where the reader can derive the important scalings.

In the limit $\Omega \gg 1$, using the parameters
$A, \psi, e, \Omega, \a$ as coordinates, the volume element is given by, 
\be
\sqrt{\det{||\tgamma_{jk}||}} = 
{A^4 \over 32 \sqrt{6}}[ e - {3 \over 4} e^3 - 
{41 \over 256} e^5 + {\cos 2 \psi \over 32}(4 e^3 - e^5 ) - {\cos 4 \psi \over 256} e^5]\,.
\label{dete}
\ee
Comparing the former expression with the numerical evaluation of the full
determinant (see Fig.~\ref{fig:ell}), it turns out -- as it happened for 
$e = 0$, cfr. the previous sections -- 
that Eq.~(\ref{dete}) is accurate within a few percent even if the number of
orbits completed during the time $T$ is just about 2 or 3. 

In order to obtain the proper volume of the parameter space, we integrate
Eq.~(\ref{dete}) over the parameter range:
\bea 
A_{\rm min} & \le & A \le A_{\rm max}\,, \nonumber\\
\Omega_{\rm min} & \le & \Omega \le \Omega_{\rm max}\,, \nonumber\\
0 & \le & e \le e_{\rm max}\,,\nonumber\\
0 & \le & \a \le 2 \pi\,,\nonumber\\
0 & \le & \psi \le 2 \pi\,,\\
\label{lime}
\eea 
and multiply by the scaling factor $\chi^5 = (2 \pi f_{\rm max} T)^5$. We
have the following expression:
\be
\label{volecc}
\Vp = {\pi^2 \over 160 \sqrt{6}}\,F(e_{\rm max})\,
\left(\frac{2\pi f_{\rm max}}{c}\right)^5\,
[\apM^5 - \apm^5]\,(\omega_{\rm max} - \omega_{\rm min})\, T\,,
\ee
where
\be
F(e) = e^2 ( 1 - {3 \over 8} e^2 - {41 \over 768} e^4)\,.
\label{ecc}
\ee
We can check how accurate this expression is by switching off the $e^6$ term 
in~(\ref{ecc}) and comparing with the full one upto $o(e^6)$. 
We find that the expressions agree very well to about $e \sim 0.7$. From this 
we surmise that the Eqs.~(\ref{volecc}) and~(\ref{ecc}) will be correct upto 
$e \sim 0.8$. We also checked how well the leading order term $e^2$ approximates
$F(e)$, and we found that up to $e = 0.5$ and $e = 0.8$ (as reference values), 
they agree within $\simeq 10\%$ and $\simeq 35\%$, respectively.

For blind searches, the size of the parameter space is:
\be
\Vp \simeq 2.3\times 10^{22}\,
\left(\frac{f_{\rm max}}{1\,{\rm kHz}}\right)^5\,
\left(\frac{\apM}{10^{11}\,{\rm cm}}\right)^5\,
\left(\frac{\omega_{\rm max}}{10^{-4}\,{\rm rad/sec}}\right)\,
\left(\frac{T}{10^{7}\,{\rm sec}}\right),
\ee
where, we have evaluated Eq.~(\ref{volecc}) for $e_{\rm max} = 0.5$. 
For $e_{\rm max} = 0.1$ and $e_{\rm max} = 0.8$ the volume changes by
the factors $\simeq 0.04$ and $\simeq 2$, respectively.
The total number of filters now is:
\be
{\cal N}(\mu = 3\%) \simeq 1.1\times 10^{26}\,\left(\frac{\Vp}{10^{22}}\right)\,.
\label{Nf_egg}
\ee
Indeed, for $e\ne 0$, prior information about the source parameters are essential
in order to reduce the computational costs, see Eq.~(\ref{power}). In particular
the steep dependence on $a_{\rm p}$ and $f$, both to the fifth power, 
indicates that the computational load could be cut down substantially if
we knew them fairly accurately.
These are the crucial parameters which govern the computational costs. ${\cal N}$
still scales linearly with $\omega T$. For sufficiently small values of the 
eccentricity, it scales quadratically with $e$.

\begin{figure}
\begin{center}
\epsfig{file=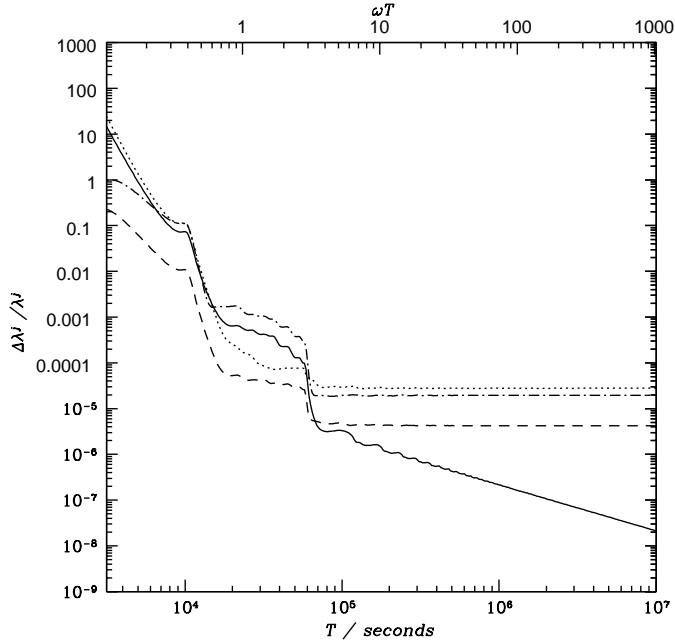,width=4.1in}
\caption{\label{fig:erre}
The errors on the orbital parameters required to carry out
a search using only a few matched filters for a binary in eccentric
orbit. The relative errors in the parameters $\Delta \omega/\omega$ (solid line), 
$\Delta X/X$ (dotted-dashed line),  
$\Delta e/e$ (dotted line) and $\Delta\alpha/\alpha$ (dashed line) are plotted 
for constant thickness equal to the filter separation $dl \simeq 0.15$,
corresponding to $\mu = 0.03$, for an edge-on eccentric orbit 
with parameters:
$\omega = 1.027\times 10^{-4}$ rad$/$sec, $a_{\rm p} = 1.64\times 10^{11}$ cm,
$e = 0.5$.
}
\end{center}
\end{figure}

As we did in Sec.~\ref{subs:circ}, 
we can estimate the thickness of the parameter space along a given direction
and compare it to the filter separation $dl$, which for $N = 5$
and $\mu = 3\%$ is $\simeq 0.15$ (see Eq.~(\ref{dl})). This is useful in obtaining 
insights on the accuracy needed on the parameters for reducing 
the computational costs by a large factor, so that they become affordable. 
In Fig.~\ref{fig:erre} we show
the relative errors $\Delta\lambda^j/\lambda^j$, 
for typical orbital elements of a NS/NS binary, as a function of the observation time,  
where we set $\cT^j = dl$.   
For observation times involving several orbits, $\Omega \gg 1$, one needs to know a 
parameter with an accuracy of the order of one part in a million, in order that the 
search is carried out with the help of only a few filters. However, as in the 
circular orbit case, 
$\Delta\lambda^j/\lambda^j$ increases very steeply when $T$ becomes of the
order of $P$, or shorter. In this case a $1\%$ accuracy (compare it
to the case $e = 0$, where $\Delta\lambda^j/\lambda^j \sim 2$) is sufficient to  
carry out the search with few filters.%
%
%
%

\section{Computational costs for known neutron stars}
\label{sec:obs}

In the previous sections we have analyzed the extra computational
costs required in correcting for the Doppler effect  
produced  by the motion of a NS around a companion object, on a 
monochromatic GW signal. We are therefore in a position to apply the 
former results to {\it known} neutron stars which are amenable to 
searches by Earth-based detectors. Here we deal with two
main classes of sources: the NS's known as radio pulsars
and the NS's in LMXB's, in particular Sco X-1.

\subsection{Radio pulsars}
\label{rp}

The last published version of the pulsar catalogue~\cite{cat} contains 
706 NS's, of which 44 are in binary systems. Actually, the total number
of pulsars known today is $\sim 1200$, thanks mainly 
to the recent multi-beam survey of the southern sky carried
out at the Parkes radio-telescope in New South Wales, 
Australia~\cite{parkes1,parkes2,parkes3}. A further considerable increase
in the number of pulsars is expected over the next few
years when the upgrade of other radio telescopes will be completed,
allowing a survey of the northern sky similar to the one carried out
at Parkes. However, the newly discovered sources have not yet been  
included into a publicly available data base, and we therefore restrict ourselves 
to the catalogue of 1995. Clearly, when the newly discovered
NS's, and their fundamental parameters will be available,
our present analysis can be applied to the new objects in a 
straightforward way~\cite{Camillo}. 

In order to compute the number of filters required to target the
44 NS's, we apply the scheme presented in
the previous sections. More specifically, we use the
full and exact expression of the parameter space volume computed 
numerically, with the values of the parameters and errors 
quoted in~\cite{cat}; the fundamental ones
are also shown in Table~\ref{tab:pulsars}. For several 
sources the orbital parameters are known very accurately,
so that we find that the search is limited either over a small region of the 
parameter space, or to a smaller number of dimensions ($< 5$), or both.  We first 
estimate the actual number of dimensions of the signal manifold, following 
the discussion in Section~\ref{subs:tar}, and then calculate the total number
of filters from Eq.~(\ref{Nf2}). We set the value of 
$\delta = 0.1$ - the thickness of the parameter space in units of
the filter distance, see Eq.~(\ref{thick2}) - which distinguishes
essentially the known parameters from the questionable ones.

\vbox{
\begin{table}[t]
\begin{center}
\caption{\label{tab:filt_rp}
The number of filters required to search for NS's known as 
radio pulsars in binaries. We list the total number of filters
${\cal N}$ required to search for the 44 NS's in binary system
contained in the Taylor et al. catalogue~{\protect\cite{cat}}
for an observation time $T = 1$ yr and a maximum mismatch of
$3\%$.  
}
\begin{tabular}{ll|c}
PSR J         & PSR B    &  ${\cal N}$           \\
\hline
 0024-7204E   & 0021-72E &  $1.8\times 10^3$ \cr
 0024-7204I   & 0021-72I &  $4.1\times 10^6$ \cr
 0024-7204J   & 0021-72J &  $9.2\times 10^2$ \cr
 0034-0534    &          &      1.           \cr
 0045-7319    &          & {\it signal outside the sensitivity window} \\
 0218+4232    &          &  $1.5\times 10^2$ \cr
 0437-4715    &          &      1            \cr
 0613-0200    &          &      1            \cr
 0700+6418    & 0655+64  &      1            \cr
 0751+1807    &          &  $1.1\times 10^7$ \cr
 0823+0159    & 0820+02  & {\it signal outside the sensitivity window} \\
 1012+5307    &          &      4            \cr
 1022+10      &          &      2            \cr
 1045-4509    &          &      16           \cr
 1300+1240    & 1257+12  &      1            \cr
 1302-6350    & 1259-63  &      2            \cr
 1312+1810    & 1310+18  &  $3.9\times 10^9$ \cr
 1455-3330    &          &      2.6          \cr
 1518+4904    &          &      2            \cr
 1537+1155    & 1534+12  &      1            \cr
 1623-2631    & 1620-26  &      8            \cr
 1640+2224    &          &      4            \cr
 1641+3627B   & 1639+36B &  $6.4\times 10^3$ \cr
 1643-1224    &          &  $7.3\times 10^4$ \cr
 1713+0747    &          &      1            \cr
 1721-1936    & 1718-19  & {\it signal outside the sensitivity window} \\
 1748-2446A   & 1744-24A &      4            \cr
 1803-2712    & 1800-27  & {\it signal outside the sensitivity window} \\
 1804-0735    & 1802-07  &      4            \cr
 1804-2718    &          &  $1.6\times 10^7$ \cr
 1823-1115    & 1820-11  & {\it signal outside the sensitivity window} \\
 1834-0010    & 1831-00  & {\it signal outside the sensitivity window} \\
 1857+0943    & 1855+09  &      1            \cr
 1910+0004    &          &  $1.2\times 10^2$ \cr
 1915+1606    & 1913+16  &      1            \cr
 1955+2908    & 1953+29  &      4            \cr
 1959+2048    & 1957+20  &      2            \cr
 2019+2425    &          &      1            \cr
 2033+17      &          &$8.9\times 10^{10}$\cr
 2130+1210C   & 2127+11C &     33            \cr
 2145-0750    &          &      4            \cr
 2229+2643    &          &     16            \cr
 2305+4707    & 2303+46  & {\it signal outside the sensitivity window} \\
 2317+1439    &          &      1            \cr
\end{tabular}
\end{center}
\end{table}%
}

The results of our analysis are presented in Table~\ref{tab:filt_rp}, 
which lists the number of templates required for an integration 
time of $T = 1$ yr. The key point to notice is that the most interesting 
NS's require at most a few filters in correcting for 
the orbital motion: J0437-4715: NS/white dwarf system; 
J0437-4715: NS/white dwarf system; B1257+12: NS/planets system;
J1537+1155: NS/NS binary; J1623-2631: NS/white dwarf system; and J1300+1240:
NS/main sequence star binary. The upper-limits on $h_{\rm c}$ are, in fact,
within the sensitivity of the first or second
generation of detectors, cfr. Figure~\ref{fig:h}. The data analysis
costs for such sources is therefore minimal. Also considering that the 
frequency is very accurately known within a small bandwidth, one can 
substantially down-sample the data stream around the expected source frequency. 
Moreover, the orbital companions of the above mentioned five neutron stars are 
also very different, which boosts the astrophysical interest in monitoring them. 

On the other hand, for those NS's about which no 
information about the spin-down is available, we cannot 
set upper-limits on the signal strain and hence the computational costs 
become considerable. The number of filters is in the range 
$\sim 10^2 - 10^{10}$, for a year's worth of integration time. Therefore, even 
assuming a search in a narrow frequency range, around twice the radio 
frequency, the computational burden is overwhelming. The situation is made 
worse by the lack of information about $\dot{f}$ and higher time derivatives 
of the frequency, so that one would also have to scan the space of spin-down 
parameters, increasing the costs further. It is clear that more accurate 
data from radio observations are necessary in order to narrow down the 
uncertainty in the source parameters for making such searches feasible.

To summarize, the interesting radio pulsars in binary systems do not
lead to large extra computational load when 
correcting for the orbital parameters of the source. On the other hand, for
the NS's for which we have no idea about the GW strain amplitude, the 
computational costs are not affordable. Such sources call for a follow-up
of radio observations for obtaining more accurate measurements of the 
orbital parameters.

\subsection{Sco X-1}
\label{sco}

Low Mass X-ray Binaries (LMXB's) could be primary GW sources for
Earth-based detectors (see discussion in the Introduction). Our current 
astrophysical understanding and signal modelling suggests that at least one 
system, namely, Sco X-1 -- the brightest X-ray source in the sky, located at 
a distance 2.8 $\pm 0.3$ kpc -- is just within the reach of the first generation 
of detectors with very long coherent observation times (of the order of
2 years), and clearly detectable by the second generation of instruments
(in particular LIGO II, in narrow-band configuration), for an integration time  
as little as $\approx 15$ days. 

Sco X-1 orbits a low-mass companion in an essentially circular orbit
with a period $P = 0.787313\, (1)$ days; the orbital angular velocity is
$\omega \simeq 9.24 \times 10^{-5}$ rad/sec and is known with
an error $\Delta\omega = 2.34 \times 10^{-10}$ rad/sec. The radial 
component of the orbital velocity is 
$v_{\rm r} = 58.2\,( 3.0)$ km/sec~\cite{GWL75}; therefore
the projection of the semi-major axis along the line of sight is 
$a_{\rm p} = v_{\rm r}/\omega\simeq 6.3\times 10^{10}\,{\rm cm}$. The
relative error on the semi-major axis is 
$\Delta a_{\rm p}/a_{\rm p} = \Delta P/P + \Delta v_{\rm r}/v_{\rm r}
\simeq \Delta v_{\rm r}/v_{\rm r}$, since the period is very accurately 
measured. This yields an error $\Delta a_{\rm p} \simeq 6.5 \times 10^9$ cm. 
The position of the NS in the orbit is known with an error 
$\Delta\alpha = 0.2$ rad~\cite{CC75}. 
The GW emission frequency is not very well determined; conservatively we assume
it to be within the band 500 Hz -- 600 Hz, and therefore we take the bandwidth 
$\Delta f \approx 100$ Hz~\cite{Bildsten98}. In the expressions for 
thicknesses of the parameter space below, it is appropriate to replace $\fM$ by 
the bandwidth $\Delta f$ over which the search would be launched.

The main problems that one faces in detecting (or setting upper limits on)
GW's emitted by Sco X-1 are: (i) the drift in the GW frequency,
which is not likely to be monotonic, and is related to the time-varying accretion 
rate  which is ultimately responsible for the net quadrupole moment, and (ii)
the need for taking into account the Doppler phase modulation caused
by the orbital motion. Present order-of-magnitude estimates of the
frequency evolution suggest that for a time duration of $\approx 10$ days the 
signal is confined within a single frequency bin; it determines the longest
coherent integration time (or base-line duration of observations) that one
can employ in searching for a monochromatic signal. As a conservative limit, we 
take a week: $T = 6.05 \times 10^5$ sec. Notice that during this
time the radiation is monochromatic, but emitted in an unknown
frequency bin inside the band 500 Hz -- 600 Hz; the position
of the source is also known very accurately, so that one is essentially 
left with the orbital parameters that one must search over. The estimate of 
the computational costs that we present therefore reflects the 
{\it total} costs involved in searching for Sco X-1.

It is interesting to consider first the thickness of the parameter
space and the conditions such that Sco X-1 is a one-filter target. 
Notice first that $\Omega = 1$ for $T = 1.08\times 10^4 {\rm sec} = 3.01$ hrs. 
In the limit $\Omega \ll 1$, therefore up to $\simeq 3$ hours,
we have:
\bea
\cT^{a_{\rm p}} & \simeq & 9.3\times 10^{-4}\,
\left(\frac{\sqrt{1 + \cos(2\alpha)}}{5 - \cos(2\alpha)}\right)\,
\left(\frac{\Delta a_{\rm p}}{6.5\times 10^{9}\,{\rm cm}}\right)\,
\left[\left(\frac{\omega}{9.24\times 10^{-5}\,{\rm sec}^{-1}}\right)\,
\left(\frac{T}{1\,{\rm hr}}\right)\right]^4\,
\left(\frac{\Delta f}{100\,{\rm Hz}}\right)\,,
\nonumber\\
\cT^{\omega} & \simeq & 1.6\times 10^{-8}\,
\left[\sqrt{1 + \cos(2\alpha)}\right]\,
\left(\frac{\Delta\omega}{2.34\times10^{-10}\,{\rm sec}^{-1}}\right)\,
\left(\frac{a_{\rm p}}{6.3\times 10^{10}\,{\rm cm}}\right)\,
\left(\frac{\omega}{9.24\times 10^{-5}\,{\rm sec}^{-1}}\right)^3\,
\left(\frac{T}{1\,{\rm hr}}\right)^4\,
\left(\frac{\Delta f}{100\,{\rm Hz}}\right)\,,
\nonumber\\
\cT^{\alpha} & \simeq & 2.6\times 10^{-3}\,
\sqrt{\frac{1 + \cos(2\alpha)}{1 - \cos(4\alpha)}}
\left(\frac{\Delta\alpha}{0.2\,{\rm rad}}\right)\,
\left(\frac{a_{\rm p}}{6.3\times 10^{10}\,{\rm cm}}\right)\,
\left[\left(\frac{\omega}{9.24\times 10^{-5}\,{\rm sec}^{-1}}\right)\,
\left(\frac{T}{1\,{\rm hr}}\right)\right]^4\,
\left(\frac{\Delta f}{100\,{\rm Hz}}\right)\,.
\label{thick_Sll}
\eea
Notice that the expressions for $\cT^{a_{\rm p}}$ and $\cT^{\omega}$ are given for
$\alpha \ne k + \pi/2$, where $k$ is an integer; the expression for $\cT^{\alpha}$ 
is valid if $\alpha \ne k\pi/2$. In the other cases one can easily derive the 
value of the thickness by applying the appropriate formula, as given in
Eq.~(\ref{thickc_ll1}). 

If one takes an hour long data stretch, Sco X-1 in a one-filter target.  
For longer observational times up to $T \approx 3$ hours, the search requires 
at most a few templates to take care of "border effects". 
A coherent search over a bandwidth $\approx 100$ Hz can be performed very inexpensively, and 
one needs a fraction of an MFlop to keep up with the data flow, which can be easily derived
from Eq.~(\ref{power}) with $\fM$ replaced by $\Delta f\approx 100$ Hz. However, 
the computational cost increases steeply (as shown in the previous 
section) when $T$ becomes of the order of the orbital period or longer. In fact, 
for $\Omega \gg 1$, the thickness of the parameter space becomes:
\bea
\cT^{a_{\rm p}} & \simeq & 96.8\,
\left(\frac{\Delta a_{\rm p}}{6.5\times 10^9\,{\rm cm}}\right)\,
\left(\frac{\Delta f}{100\,{\rm Hz}}\right)\,,
\nonumber\\
\cT^{\omega} & \simeq & 6.3\times 10^{-3}\,
\left(\frac{\Delta\omega}{2.3 \times 10^{-10}\,{\rm sec}^{-1}}\right)\,
\left(\frac{a_{\rm p}}{6.3\times 10^{10}\,{\rm cm}}\right)\,
\left(\frac{T}{10^5\,{\rm sec}}\right)\,
\left(\frac{\Delta f}{100\,{\rm Hz}}\right)\,,
\nonumber\\
\cT^{\alpha} & \simeq & 93.6\,
\left(\frac{\Delta\alpha}{0.2\,{\rm rad}}\right)\,
\left(\frac{a_{\rm p}}{6.3\times 10^{10}\,{\rm cm}}\right)\,
\left(\frac{\Delta f}{100\,{\rm Hz}}\right)\,.
\label{thick_Sgg}
\eea
One needs a mesh of filters for $a_{\rm p}$ and $\a$, but up to
$T \approx 1$ month the orbital period is not a search parameter
(one might actually need up to 2 filters to take care of the
errors in $P$). The reduced metric $\bar{\gamma}_{jk}$ which 
we need to consider is given by a $2\times 2$ matrix on $a_{\rm p}$ and $\a$.
By applying Eq.~(\ref{Nf1}),  with $\mu = 0.03$ and $T \simlt$ 1 month, we obtain,
\be
{\cal N}_{(N = 2)} \simeq 5.2\times 10^6\,
\left[1 - \left(\frac{\delta^2}{3}\right)\right]^{-1}\,.
\label{Nsco1}
\ee
A one week coherent search involving $5\times 10^6$ filters requires $\approx 84$
GFlops of processing power to keep up with the data flow.
For longer integration times $T \simgt$ 1 month, $\omega$ becomes a search 
parameter; notice that in this case the signal is not
anymore monochromatic, and one needs to search also over the spin-downs 
parameter space (which is not trivial, due to our poor understanding of the
accretion rate and angular momentum transfer, see Sec.~\ref{subs:sm});
the number of filters to correct {\it only} for the Doppler shift induced by Sco X-1
orbital motion is given now by:
\be
{\cal N}_{(N = 3)} \simeq 3.1 \times 10^8\, 
\left(\frac{T}{10^7\,{\rm sec}}\right) \,.
\label{Nsco2}
\ee
By setting a coarser grid, increasing the mismatch from, say, $3\%$ to $30 \%$,
will reduce the number of filters in the two cases~(\ref{Nsco1}) and~(\ref{Nsco2}) 
by a factor $\simeq 10$ and $\simeq 30$, respectively. The values still indicate
that the computational costs are too high which is also due to the large 
frequency band that one needs to search through. These results clearly show that
it is mandatory to determine the orbital parameters more accurately in order to 
reduce the parameter volume so that Sco X-1 can be monitored continuously. 
Eq.~(\ref{thick_Sgg}) implies that the errors
on $a_{\rm p}$ and $\a$ must be reduced by 3 orders-of-magnitude, so that only 
a few templates are required for the search. Although this calls for 
a very large observational step, it is still possible.

%
%

\section{Conclusions}
\label{sec:concl}

We have estimated the extra-computational costs 
required to search for CW sources in binary systems. This work
was motivated by the fact that LMXB's, a key class of sources, possibly, 
for the initial generation of detectors and certainly for the enhanced detectors,  
are NS's orbiting a binary companion. Also several NS's detected in the radio band  
are in binaries and will be continously monitored by the Earth-based network of 
GW detectors. In order to disentangle the computational costs arising from 
correcting for the Doppler phase modulation produced by the source motion
around the orbital companion, we have assumed that the signal is exactly
monochromatic and the location of the source in the sky is perfectly known. 
Our results provide an estimate of the computational
burden associated with these extra dimensions of the parameter space which has 
been neglected so far (for rather obvious reasons). 
We have deduced closed form analytical expressions for the
number of templates required to carry out the analysis. We have obtained the 
{\it scaling} of the computational costs as a function of the orbital elements and 
the observational parameters (the source emission
frequency and the observation time); they are ready-to-use tools to 
evaluate the computational resources needed and the trade off for search strategies. 
We have applied our analysis to known radio pulsars and Sco X-1, estimating
the computational costs involved in targeting these systems. The analysis 
also addresses the need of further astronomical observations
of known NS's, with the aim of reducing the errors in the source parameters, 
so that  
the search is made inexpensive. In fact, a byproduct of our analysis is 
the rigorous formulation, in the geometrical framework of data analysis, 
of the conditions, {\it i.e.} the size of the parameter space, such
that a NS can be considered a one-filter target; our result is completely 
general, and can be applied to any class of sources (not necessarily CW sources) 
whose waveforms can be modelled reliably.

Our analysis is clearly limited in several respects. Possibly, the main
accomplishment is a quantitative understanding of the
key issues that require intensive attack, and identifying the bottlenecks
that prevent us from carrying out blind searches of NS in binary systems. 
There are three clear directions of future work that
we believe should be pursued vigorously, particularly in the light of the current 
data analysis effort in setting up GW search codes: 

\begin{enumerate}

\item The software implementation of matched filter based searches of a NS in a 
binary system, for given values of the source parameters such as location in the 
sky, spin-down(s), orbital elements. 
This task is simple in principle,  as the fundamental
building blocks are already available: search codes for 
GW's from isolated NS's and publicly available software (such as TEMPO) 
in the radio pulsar community to correct for the Doppler effect of the signal 
phase for a source in a binary system. However, the implementation of this might 
not prove to be so trivial.

\item The estimation of the computational costs keeping into account
{\it both} the orbital elements {\it and} the the source position
and the spin-down parameters. This is the composite problem where the full 
expression for the total costs needs to be obtained. It will take into account 
possible correlations between the search parameters investigated here, and those 
that describe isolated neutron stars.

\item The investigation of suitable {\it hierarchical strategies}, that
can reduce the computational costs as compared to fully coherent
searches. The current vigorous effort in the study, and software implementation, 
of hierarchical algorithms for isolated sources, the so-called stack-and-slide
and Hough-pattern-tracking searches provide already a template for
investigations for a larger parameter space. Moreover, radio and X-ray
astronomers are currently investigating algorithms suitable to carry out
analogous searches in the electro-magnetic spectrum~\cite{JK91,Ransom00}.

\end{enumerate}

One limitation of our approach is that we have 
completely neglected relativistic effects in the model of a binary
orbit. In particular, the advance of the periapse
is very significant in relativistic systems (the largest periapse
advance reported in~\cite{cat} is $\approx 4$ deg/yr). This, and
other complications, with respect to the simple Keplerian model that we
have adopted here, need to be taken into account in future studies,
and addressed carefully. 

We would like to conclude by stressing that the data analysis problem that we have 
addressed here is common to electro-magnetic observations 
of pulsating stars in binary orbits, where typical examples are
radio and X-rays binaries. We can therefore hope that the current efforts in other 
related research fields can help in providing a breakthrough in this area.

\acknowledgements

We would like to thank C. Cutler, M.~A. Papa and B. Schutz for helpful discussions.

\appendix

\section{Coefficients of the Taylor expansion of the eccentric anomaly}
\label{app:CS}

Here we give the expressions of the coefficients $\Ck(e)$ and $\Sk(e)$ 
which appear in the expressions for $\cos E$ and $\sqrt{1 - e^2} \sin E$ in 
Eq.~(\ref{pwrser}), upto the 7-th order in $e$~\cite{Taff}:
\bea
{\cal C}_0 &=& - \h e \,,\nonumber\\
{\cal C}_1 &=& 1 - {3 \over 8} e^2 + {5 \over 192} e^4 - {7 \over 9216} e^6 \,,\nonumber\\
{\cal C}_2  &=& \h e - {1 \over 3} e^3 + {1 \over 16}e^5 \,,\nonumber\\
{\cal C}_3  &=& {3 \over 8} e^2 - {45 \over 128} e^4 +{567 \over 5120} e^6 \,,\nonumber\\
{\cal C}_4  &=& {1 \over 3} e^3 - {2 \over 5} e^5 \,,\nonumber\\
{\cal C}_5  &=& {125 \over 384} e^4 - {4375 \over 9216} e^6 \,,\nonumber\\
{\cal C}_6  &=& {27 \over 80} e^5 \,,\nonumber\\
{\cal C}_7  &=& {16807 \over 46080} e^6\,;
\eea
\bea
{\cal S}_1 &=& 1 - {5 \over 8} e^2 - {11 \over 192} e^4 - {457 \over 9216} e^6 \,,\nonumber\\
{\cal S}_2 &=& \h e - {5 \over 12} e^3 + {1 \over 24} e^5 \,,\nonumber\\
{\cal S}_3 &=& {3 \over 8}e^2 - {51 \over 128} e^4 + {543 \over 5120} e^6,\,,\nonumber\\
{\cal S}_4 &=& {1 \over 3} e^3 - {13 \over 30} e^5 \,,\nonumber\\
{\cal S}_5 &=& {125 \over 384} e^4 - {4625 \over 9216} e^6 \,,\nonumber\\
{\cal S}_6 &=& {27 \over 80} e^5 \,,\nonumber\\
{\cal S}_7 &=& {16807 \over 46080} e^6 \,;\nonumber\\
\eea
we refer the reader to~\cite{Taff}, or any other standard book of celestial mechanics
for the general expressions of the coefficients $\Ck(e)$ and $\Sk(e)$.

\section{The approximate metric components for circular orbits}
\label{app:e0}

We provide the the expressions, to the relevant order in $A$ and $\Omega$, 
of the components of the metrics
$\tg_{\a \b}$ and $\tgamma_{jk}$, see Eqs.~(\ref{metric}),~(\ref{Xav}),~(\ref{gamma})
and~(\ref{sc_g}), for a NS in circular
orbit, using the phase model~(\ref{circphs}) in the two asymptotic
limits $\Omega \gg 1$ and $\Omega \ll 1$. 

\subsection{The asymptotic limit of several orbits}
\label{app:e0gg1}

We derive the expressions of the components of $\tg_{\a \b}$ ($\a,\b = 0,1,2,3$),
and $\tgamma_{jk}$ ($i,j = 1,2,3$) given by Eq.~(\ref{sc_g}) for
$\Omega \gg 1$. We recall that, in general, $A\ll 1/\Omega$.
The time derivatives are given in Eq.~(\ref{circder}), and
the time averages are computed through Eq.~(\ref{Xav}).

To the leading order in $A$ and $\Omega$ the elements of $\tg_{\a \b}$ read:
\bea
\tg_{00} & = & {\frac{1}{12}}\,, \\
\tg_{01} & = &   -{\frac{\cos\a}{{{\Omega}^2}}} + 
   {\frac{\cos (\a + \Omega)}{{{\Omega}^2}}}\,,  \\
\tg_{02} & = & {\frac{A\,\cos (\a + \Omega)}{2\,\Omega}}\,,  \\
\tg_{03} & = &   {\frac{A\,\cos\a}{2\,\Omega}} + {\frac{A\,\cos (\a + \Omega)}{2\,\Omega}}\,,  \\
\tg_{11} & = &   {\frac{1}{2}}  - 
   {\frac{\sin (2\a)}{4\,\Omega}} + 
   {\frac{\sin [2\,\left( a + \Omega \right) ]}{4\,\Omega}}\,, \\
\tg_{12} & = &   {\frac{A\,\cos [2(\a + \Omega)]}{4\,\Omega}}\,, \\
\tg_{13} & = &   -{\frac{A\,{\cos^2\a}}{2\,\Omega}} + 
   {\frac{A\,{{\cos^2 (\a + \Omega)}}}{2\,\Omega}}\,, \\
\tg_{22} & = &   {\frac{{A^2}}{6}}\,,  \\
\tg_{23} & = &   {\frac{{A^2}}{4}}\,, \\
\tg_{33} & = &   {\frac{{A^2}}{2}}\,. 
\eea
Substituting these components into Eq.~(\ref{gamma}), which is
related to $\tgamma_{jk}$ through Eq.~(\ref{sc_g}), one derives the following
expressions, to the leading order in $A$ and $\Omega$:
\bea
\tgamma_{11} & = & {\frac{1}{2}} - {\frac{\sin (2\a)}{4\,\Omega}} + 
   {\frac{\sin [2\,\left( a + \Omega \right) ]}{4\,\Omega}}\,, \\
\tgamma_{12} & = & 
   {\frac{A\,\cos [2(\a + \Omega)]}{4\,\Omega}} \,, \\
\tgamma_{13} & = & 
   - {\frac{A\,{{\cos\a}^2}}{2\,\Omega}}  + 
   {\frac{A\,\cos^2 (\a + \Omega)}{2\,\Omega}}\,, \\
\tgamma_{22} & = &
  {\frac{{A^2}}{6}}\,, \\
\tgamma_{23} & = &
 {\frac{{A^2}}{4}}\,,  \\
\tgamma_{33} & = &
  {\frac{{A^2}}{2}} \,.
\eea
The determinant~(\ref{detc1}) is obtained from the latter components by 
retaining the leading order terms in $A$ and $1/\Omega$.

\subsection{The asymptotic limit of a fraction of one orbit}
\label{app:e0ll1}

The expression of the determinant in the case $\Omega \ll 1$
can be simply derived by Taylor expanding the components of 
$\tg_{\a \b}$ and $\tgamma_{jk}$ in powers of $\Omega$. Here
we present the expressions of $\tg_{\a \b}$ upto $\Omega^{10}$:
\bea
\tg_{00} & = & {\frac{1}{12}} \\
\tg_{01} & = & 
  {\frac{-\left( {\Omega^2}\,\cos \a \right) }{24}} + 
   {\frac{{\Omega^4}\,\cos \a}{360}} - {\frac{{\Omega^6}\,\cos \a}{13440}} + 
   {\frac{{\Omega^8}\,\cos \a}{907200}} - {\frac{{\Omega^{10}}\,\cos \a}{95800320}} - 
   {\frac{\Omega\,\sin \a}{12}} + {\frac{{\Omega^3}\,\sin \a}{80}} 
\nonumber \\     &&
 - {\frac{{\Omega^5}\,\sin \a}{2016}} + {\frac{{\Omega^7}\,\sin \a}{103680}} - 
   {\frac{{\Omega^9}\,\sin \a}{8870400}}
   \\
\tg_{02} & = &
  {\frac{-\left( A\,\Omega\,\cos \a \right) }{12}} + 
   {\frac{A\,{\Omega^3}\,\cos \a}{90}} - {\frac{A\,{\Omega^5}\,\cos \a}{2240}} + 
   {\frac{A\,{\Omega^7}\,\cos \a}{113400}} - 
   {\frac{A\,{\Omega^9}\,\cos \a}{9580032}} - {\frac{A\,\sin \a}{12}} + 
\nonumber \\     &&
   {\frac{3\,A\,{\Omega^2}\,\sin \a}{80}} - 
   {\frac{5\,A\,{\Omega^4}\,\sin \a}{2016}} + 
   {\frac{7\,A\,{\Omega^6}\,\sin \a}{103680}} - 
   {\frac{A\,{\Omega^8}\,\sin \a}{985600}} + 
   {\frac{11\,A\,{\Omega^{10}}\,\sin \a}{1132185600}}
   \\
\tg_{03} & = &
  {\frac{-\left( A\,\Omega\,\cos \a \right) }{12}} + 
   {\frac{A\,{\Omega^3}\,\cos \a}{80}} - {\frac{A\,{\Omega^5}\,\cos \a}{2016}} + 
   {\frac{A\,{\Omega^7}\,\cos \a}{103680}} - 
   {\frac{A\,{\Omega^9}\,\cos \a}{8870400}} + {\frac{A\,{\Omega^2}\,\sin \a}{24}} - 
\nonumber \\     &&
   {\frac{A\,{\Omega^4}\,\sin \a}{360}} + {\frac{A\,{\Omega^6}\,\sin \a}{13440}} - 
   {\frac{A\,{\Omega^8}\,\sin \a}{907200}} + 
   {\frac{A\,{\Omega^{10}}\,\sin \a}{95800320}}
   \\
\tg_{11} & = &
  {\Omega^2}\,\left( {\frac{1}{24}} - {\frac{\cos (2\a)}{24}} \right)  + 
   {\Omega^6}\,\left( {\frac{1}{40320}} - {\frac{43\,\cos (2\a)}{13440}} \right)
       + {\Omega^{10}}\,\left( {\frac{1}{479001600}} - 
      {\frac{4097\,\cos (2\a)}{479001600}} \right)  + 
\nonumber \\     &&
   {\Omega^8}\,\left( -{\frac{1}{3628800}} + 
      {\frac{769\,\cos (2\a)}{3628800}} \right)  + 
   {\Omega^4}\,\left( -{\frac{1}{720}} + {\frac{17\,\cos (2\a)}{720}} \right)  + 
   {\frac{{\Omega^3}\,\sin (2\a)}{24}} - {\frac{7\,{\Omega^5}\,\sin (2\a)}{720}} + 
\nonumber \\     &&
   {\frac{107\,{\Omega^7}\,\sin (2\a)}{120960}} - 
   {\frac{163\,{\Omega^9}\,\sin (2\a)}{3628800}}
   \\
\tg_{12} & = &
  \Omega\,\left( {\frac{A}{24}} - {\frac{A\,\cos (2\a)}{24}} \right)  + 
   {\Omega^5}\,\left( {\frac{A}{13440}} - {\frac{43\,A\,\cos (2\a)}{4480}} \right)
       + {\Omega^9}\,\left( {\frac{A}{95800320}} - 
      {\frac{4097\,A\,\cos (2\a)}{95800320}} \right)  + 
\nonumber \\     &&
   {\Omega^7}\,\left( {\frac{-A}{907200}} + 
      {\frac{769\,A\,\cos (2\a)}{907200}} \right)  + 
   {\Omega^3}\,\left( {\frac{-A}{360}} + {\frac{17\,A\,\cos (2\a)}{360}} \right)
       + {\frac{A\,{\Omega^2}\,\sin (2\a)}{16}} - 
\nonumber \\     &&
   {\frac{7\,A\,{\Omega^4}\,\sin (2\a)}{288}} + 
   {\frac{107\,A\,{\Omega^6}\,\sin (2\a)}{34560}} - 
   {\frac{163\,A\,{\Omega^8}\,\sin (2\a)}{806400}} + 
   {\frac{709\,A\,{\Omega^{10}}\,\sin (2\a)}{87091200}}
   \\
\tg_{13} & = &
  {\frac{A\,{\Omega^3}\,\cos (2\a)}{24}} - 
   {\frac{7\,A\,{\Omega^5}\,\cos (2\a)}{720}} + 
   {\frac{107\,A\,{\Omega^7}\,\cos (2\a)}{120960}} - 
   {\frac{163\,A\,{\Omega^9}\,\cos (2\a)}{3628800}} + 
   {\frac{A\,{\Omega^2}\,\sin (2\a)}{24}} - 
\nonumber \\     &&
   {\frac{17\,A\,{\Omega^4}\,\sin (2\a)}{720}} + 
   {\frac{43\,A\,{\Omega^6}\,\sin (2\a)}{13440}} - 
   {\frac{769\,A\,{\Omega^8}\,\sin (2\a)}{3628800}} + 
   {\frac{4097\,A\,{\Omega^{10}}\,\sin (2\a)}{479001600}}
   \\
\tg_{22} & = &
  {\frac{{A^2}}{24}} - {\frac{{A^2}\,\cos (2\a)}{24}} + 
   {\Omega^4}\,\left( {\frac{-{A^2}}{5760}} - 
      {\frac{113\,{A^2}\,\cos (2\a)}{4480}} \right)  + 
   {\Omega^8}\,\left( {\frac{-{A^2}}{43545600}} - 
      {\frac{93173\,{A^2}\,\cos (2\a)}{479001600}} \right)  + 
\nonumber \\     &&
   {\Omega^{10}}\,\left( {\frac{{A^2}}{6706022400}} + 
      {\frac{75547\,{A^2}\,\cos (2\a)}{9686476800}} \right)  + 
   {\Omega^6}\,\left( {\frac{{A^2}}{403200}} + 
      {\frac{10999\,{A^2}\,\cos (2\a)}{3628800}} \right)  + 
\nonumber \\     &&
   {\Omega^2}\,\left( {\frac{{A^2}}{144}} + 
      {\frac{59\,{A^2}\,\cos (2\a)}{720}} \right)  + 
   {\frac{{A^2}\,\Omega\,\sin (2\a)}{12}} - 
   {\frac{19\,{A^2}\,{\Omega^3}\,\sin (2\a)}{360}} + 
   {\frac{29\,{A^2}\,{\Omega^5}\,\sin (2\a)}{3024}} - 
\nonumber \\     &&
   {\frac{1489\,{A^2}\,{\Omega^7}\,\sin (2\a)}{1814400}} + 
   {\frac{4913\,{A^2}\,{\Omega^9}\,\sin (2\a)}{119750400}}
   \\
\tg_{23} & = &
  {\Omega^4}\,\left( {\frac{-{A^2}}{1440}} - 
      {\frac{7\,{A^2}\,\cos (2\a)}{288}} \right)  + 
   {\Omega^8}\,\left( {\frac{-{A^2}}{7257600}} - 
      {\frac{163\,{A^2}\,\cos (2\a)}{806400}} \right)  + 
\nonumber \\     &&
   {\Omega^{10}}\,\left( {\frac{{A^2}}{958003200}} + 
      {\frac{709\,{A^2}\,\cos (2\a)}{87091200}} \right)  + 
   {\Omega^6}\,\left( {\frac{{A^2}}{80640}} + 
      {\frac{107\,{A^2}\,\cos (2\a)}{34560}} \right)  + 
   {\Omega^2}\,\left( {\frac{{A^2}}{48}} + {\frac{{A^2}\,\cos (2\a)}{16}} \right)
\nonumber \\     &&
       + {\frac{{A^2}\,\Omega\,\sin (2\a)}{24}} - 
   {\frac{17\,{A^2}\,{\Omega^3}\,\sin (2\a)}{360}} + 
   {\frac{43\,{A^2}\,{\Omega^5}\,\sin (2\a)}{4480}} - 
   {\frac{769\,{A^2}\,{\Omega^7}\,\sin (2\a)}{907200}} + 
   {\frac{4097\,{A^2}\,{\Omega^9}\,\sin (2\a)}{95800320}}
   \\
\tg_{33} & = &
  {\Omega^4}\,\left( {\frac{-{A^2}}{720}} - 
      {\frac{17\,{A^2}\,\cos (2\a)}{720}} \right)  + 
   {\Omega^8}\,\left( {\frac{-{A^2}}{3628800}} - 
      {\frac{769\,{A^2}\,\cos (2\a)}{3628800}} \right)  + 
   {\Omega^6}\,\left( {\frac{{A^2}}{40320}} + 
      {\frac{43\,{A^2}\,\cos (2\a)}{13440}} \right)  + 
\nonumber \\     &&
   {\Omega^{10}}\,\left( {\frac{{A^2}}{479001600}} + 
      {\frac{4097\,{A^2}\,\cos (2\a)}{479001600}} \right)  + 
   {\Omega^2}\,\left( {\frac{{A^2}}{24}} + {\frac{{A^2}\,\cos (2\a)}{24}} \right)
       - {\frac{{A^2}\,{\Omega^3}\,\sin (2\a)}{24}} + 
\nonumber \\     &&
   {\frac{7\,{A^2}\,{\Omega^5}\,\sin (2\a)}{720}} - 
   {\frac{107\,{A^2}\,{\Omega^7}\,\sin (2\a)}{120960}} + 
   {\frac{163\,{A^2}\,{\Omega^9}\,\sin (2\a)}{3628800}}
\eea
Substituting the former components into Eq.~(\ref{gamma}), which is
related to $\tgamma_{jk}$ through Eq.~(\ref{sc_g}), one derives the following
expressions for the components of the metric $\tgamma_{jk}$ upto order $\Omega^{10}$:
\bea
\tgamma_{11} & = &
  {\Omega^6}\,\left( -{\frac{1}{50400}} - {\frac{19\,\cos (2\a)}{50400}} \right)
       + {\Omega^{10}}\,\left( -{\frac{1}{419126400}} - 
      {\frac{521\,\cos (2\a)}{209563200}} \right)  + 
   {\Omega^8}\,\left( {\frac{1}{3628800}} + {\frac{23\,\cos (2\a)}{518400}}
      \right)  
\nonumber \\     &&      
      + {\Omega^4}\,\left( {\frac{1}{1440}} + 
      {\frac{\cos (2\a)}{1440}} \right)  - 
   {\frac{{\Omega^5}\,\sin (2\a)}{1440}} + 
   {\frac{11\,{\Omega^7}\,\sin (2\a)}{75600}} - 
   {\frac{41\,{\Omega^9}\,\sin (2\a)}{3628800}}
   \\
\tgamma_{12} & = &
  {\Omega^5}\,\left( {\frac{-A}{16800}} - {\frac{19\,A\,\cos (2\a)}{16800}}
      \right)  + {\Omega^9}\,\left( {\frac{-A}{83825280}} - 
      {\frac{521\,A\,\cos (2\a)}{41912640}} \right)  + 
   {\Omega^7}\,\left( {\frac{A}{907200}} + {\frac{23\,A\,\cos (2\a)}{129600}}
       \right)  + 
\nonumber \\     &&       
       {\Omega^3}\,\left( {\frac{A}{720}} + 
      {\frac{A\,\cos (2\a)}{720}} \right)  - 
   {\frac{A\,{\Omega^4}\,\sin (2\a)}{576}} + 
   {\frac{11\,A\,{\Omega^6}\,\sin (2\a)}{21600}} - 
   {\frac{41\,A\,{\Omega^8}\,\sin (2\a)}{806400}} + 
   {\frac{2027\,A\,{\Omega^{10}}\,\sin (2\a)}{762048000}}
   \\
\tgamma_{13} & = &
  {\frac{-\left( A\,{\Omega^5}\,\cos (2\a) \right) }{1440}} + 
   {\frac{11\,A\,{\Omega^7}\,\cos (2\a)}{75600}} - 
   {\frac{41\,A\,{\Omega^9}\,\cos (2\a)}{3628800}} - 
   {\frac{A\,{\Omega^4}\,\sin (2\a)}{1440}} + 
   {\frac{19\,A\,{\Omega^6}\,\sin (2\a)}{50400}} - 
\nonumber \\     &&       
   {\frac{23\,A\,{\Omega^8}\,\sin (2\a)}{518400}} + 
   {\frac{521\,A\,{\Omega^{10}}\,\sin (2\a)}{209563200}}
   \\
\tgamma_{22} & = & 
  {\Omega^4}\,\left( {\frac{{A^2}}{50400}} - 
      {\frac{23\,{A^2}\,\cos (2\a)}{7200}} \right)  + 
   {\Omega^8}\,\left( {\frac{13\,{A^2}}{838252800}} - 
      {\frac{48509\,{A^2}\,\cos (2\a)}{838252800}} \right)  + 
   {\Omega^{10}}\,\left( {\frac{-{A^2}}{7925299200}} + 
      {\frac{413747\,{A^2}\,\cos (2\a)}{145297152000}} \right)  
\nonumber \\     &&      
      + {\Omega^6}\,\left( {\frac{-{A^2}}{907200}} + 
      {\frac{599\,{A^2}\,\cos (2\a)}{907200}} \right)  + 
   {\Omega^2}\,\left( {\frac{{A^2}}{360}} + {\frac{{A^2}\,\cos (2\a)}{360}}
      \right)  - {\frac{{A^2}\,{\Omega^3}\,\sin (2\a)}{240}} + 
\nonumber \\     &&
   {\frac{503\,{A^2}\,{\Omega^5}\,\sin (2\a)}{302400}} - 
   {\frac{193\,{A^2}\,{\Omega^7}\,\sin (2\a)}{907200}} + 
   {\frac{114431\,{A^2}\,{\Omega^9}\,\sin (2\a)}{8382528000}}
   \\
\tgamma_{23} & = &
  {\Omega^4}\,\left( {\frac{{A^2}}{2880}} - {\frac{{A^2}\,\cos (2\a)}{576}}
      \right)  + {\Omega^8}\,\left( {\frac{{A^2}}{7257600}} - 
      {\frac{41\,{A^2}\,\cos (2\a)}{806400}} \right)  + 
   {\Omega^{10}}\,\left( {\frac{-{A^2}}{838252800}} + 
      {\frac{2027\,{A^2}\,\cos (2\a)}{762048000}} \right)  + 
\nonumber \\     &&
   {\Omega^6}\,\left( {\frac{-{A^2}}{100800}} + 
      {\frac{11\,{A^2}\,\cos (2\a)}{21600}} \right)  - 
   {\frac{{A^2}\,{\Omega^3}\,\sin (2\a)}{720}} + 
   {\frac{19\,{A^2}\,{\Omega^5}\,\sin (2\a)}{16800}} - 
   {\frac{23\,{A^2}\,{\Omega^7}\,\sin (2\a)}{129600}} + 
   {\frac{521\,{A^2}\,{\Omega^9}\,\sin (2\a)}{41912640}}
   \\
\tgamma_{33} & = &
  {\Omega^4}\,\left( {\frac{{A^2}}{1440}} - {\frac{{A^2}\,\cos (2\a)}{1440}}
       \right)  + {\Omega^8}\,\left( {\frac{{A^2}}{3628800}} - 
      {\frac{23\,{A^2}\,\cos (2\a)}{518400}} \right)  + 
   {\Omega^{10}}\,\left( {\frac{-{A^2}}{419126400}} + 
      {\frac{521\,{A^2}\,\cos (2\a)}{209563200}} \right)  + 
\nonumber \\     &&
   {\Omega^6}\,\left( {\frac{-{A^2}}{50400}} + 
      {\frac{19\,{A^2}\,\cos (2\a)}{50400}} \right)  + 
   {\frac{{A^2}\,{\Omega^5}\,\sin (2\a)}{1440}} - 
   {\frac{11\,{A^2}\,{\Omega^7}\,\sin (2\a)}{75600}} + 
   {\frac{41\,{A^2}\,{\Omega^9}\,\sin (2\a)}{3628800}}
\eea
The determinant~(\ref{detc2}) is then derived from the 
latter components and Taylor expanding the resulting expression to 
the leading order in $\Omega$.

\section{The approximate metric components for elliptic orbits}
\label{app:ell1}

Below we list the metric components $\tg_{\a \b}$ for $\a \ge \b$, in the 
coordinates $\kappa, X, Y, e, \psi, \a$ upto order $e^2$ and in the leading order
$X, Y$ for the case $\Omega \gg 1$: 
\bea
\tg_{00} &=& {1 \over 12}, \\
\tg_{0X} &=& - {1 \over 4} e^2 X, \\ 
\tg_{0Y} &=& 0, \\
\tg_{0e} &=& - {1 \over 4} e X^2, \\
\tg_{0 \Omega} &=& 0, \\
\tg_{0 \a}  &=& 0, \\
\tg_{XX} &=& \h (1 - \h e^2), \\
\tg_{XY}  &=& 0, \\
\tg_{Xe} &=& - {1 \over 4} e X, \\
\tg_{X \Omega}  &=&  {1 \over 4} (1 - \h e^2) Y, \\
\tg_{X \a} &=& \h (1 - \h e^2) Y, \\
\tg_{YY} &=& \h (1 - e^2), \\
\tg_{Ye} &=& - \h e Y, \\
\tg_{Y \Omega} &=& - {1 \over 4}(1 - \h e^2) X, \\
\tg_{Y \a}  &=& - \h (1 - \h e^2) X, \\
\tg_{ee} &=& {1 \over 8}(X^2 + Y^2) +  {e^2 \over 16} (X^2 + 7 Y^2),\\
\tg_{e \Omega} &=& {1 \over 8} e X Y, \\
\tg_{e \a} &=& {1 \over 4} e X Y, \\
\tg_{\Omega \Omega} &=& {1 \over 6} (X^2 + Y^2) + {e^2 \over 24} (X^2 - Y^2), \\
\tg_{\Omega \a} &=& {1 \over 4} (X^2 + Y^2) + {e^2 \over 16} (X^2 - Y^2), \\
\tg_{\a \a} &=&  {1 \over 2} (X^2 + Y^2) + {e^2 \over 8} (X^2 - Y^2)\,. 
\eea
In the previous expressions we have also dropped all the terms of order 
$\Omega^{-1}$ and higher, and set $\alpha = 0$. Substituting these components
$\tg_{jk}$ into Eq.~(\ref{gamma}), and consistently working within the
appropriate approximations, one can derive (to the leading order in $e^2$),
the expression for the determinant in ~(\ref{dete}).

%
%
%
%

\end{document}